 \documentstyle[aas2pp4,epsf]{article}

\newcommand{\Ell}{E_\parallel}      
\newcommand{\rhoGJ}{\rho_{{\rm GJ}}}  
\newcommand{\sgT}{\sigma_{\rm T}}  
\newcommand{\sgP}{\sigma_{\rm p}}  
\newcommand{\rlc}{\varpi_{\rm LC}} 
\newcommand{\Ex}{\epsilon_{\rm x}} 
\newcommand{\Eg}{\epsilon_\gamma}  
\newcommand{\inc}{\alpha_{\rm i}}  

\begin{document}

\title{GAMMA-RAY EMISSIONS FROM PULSARS:
       SPECTRA OF THE TEV FLUXES FROM OUTER-GAP ACCELERATORS}
\author{K. Hirotani}
\affil{National Astronomical Observatory, 
       Mitaka, Tokyo 181-8588, Japan;
       hirotani@hotaka.mtk.nao.ac.jp}
\authoremail{hirotani@hotaka.mtk.nao.ac.jp}

\begin{abstract}
We study the gamma-ray emissions from an outer-magnetospheric potential gap
around a rotating neutron star.
Migratory electrons and positrons are accelerated by the
electric field in the gap to radiate copious gamma-rays
via curvature process.
Some of these gamma-rays materialize as pairs
by colliding with the X-rays in the gap, 
leading to a pair production cascade. 
Imposing the closure condition that a single pair produces
one pair in the gap on average,  
we explicitly solve the strength of the acceleration field 
and demonstrate how the peak energy and the luminosity
of the curvature-radiated, GeV photons 
depend on the strength of the surface blackbody and the power-law
emissions. 
Some predictions on the GeV emission from
twelve rotation-powered pulsars are presented.
We further demonstrate that the expected pulsed TeV fluxes
are consistent with their observational upper limits.
An implication of high-energy pulse phase width versus
pulsar age, spin, and magnetic moment is discussed.
\end{abstract}



\section{Introduction}

The EGRET experiment on the Compton Gamma Ray Observatory
has detected pulsed signals from seven rotation-powered pulsars 
(e.g., Nolan et al. 1996, and references therein;
Kapsi et al. 2000):
Crab, Vela, Geminga, PSR B1706-44, PSR B1951+32, PSR B1046-58,
and PSR B1055-52, 
with PSR B0656+14 being a possible detection
(Ramanamurthy et al. 1996).
The modulation of the $\gamma$-ray light curves at GeV energies 
testifies to the production of $\gamma$-ray radiation in the pulsar 
magnetospheres either at the polar cap 
(Harding, Tademaru, \& Esposito 1978; Daugherty \& Harding 1982, 1996;
 Dermer \& Sturner 1994; Sturner, Dermer, \& Michel 1995;
 Shibata, Miyazaki, \& Takahara 1998; Miyazaki \& Takahara 1997;
 also see Scharlemann, Arons, \& Fawley 1978 for the slot gap model),
or at the vacuum gaps in the outer magnetosphere
(Chen, Ho, \& Ruderman 1986a,b, hereafter CHR;
 Chiang \& Romani 1992, 1994; Romani and Yadigaroglu 1995;
 Romani 1996; Zhang \& Cheng 1998; Cheng \& Zhang 1999).
Effective $\gamma$-ray production in a pulsar magnetosphere
may be extended to the very high energy (VHE) region above 
100 GeV as well;
however, the predictions of fluxes by the current models of 
$\gamma$-ray pulsars are not sufficiently conclusive
(e.g., Cheng 1994).
Whether or not the spectra of $\gamma$-ray pulsars continue up to the
VHE region is a question which remains one of the 
interesting issues of high-energy astrophysics.

In the VHE region,
positive detections of radiation at a high confidence 
level have been reported from the direction of the 
Crab, B1706-44, and Vela pulsars 
(Bowden et al. 1993; Nel et al. 1993; Edwards et al. 1994;
 Yoshikoshi et al. 1997; 
 see also Kifune 1996 for a review),
by virtue of the technique of imaging Cerenkov light
from extensive air showers.
However, with respect to {\it pulsed} TeV emissions,
only the upper limits have been, as a rule, obtained from these pulsars
(see the references cited just above). 
If the VHE emission originates the pulsar magnetosphere,
rather than the extended nebula,
significant fraction of them can be expected to show a pulsation.
Therefore, the lack of {\it pulsed} TeV emissions provides a
severe constraint on the modeling of particle acceleration zones
in a pulsar magnetosphere.

In fact, in CHR picture,
the magnetosphere should be optically thick for pair production
in order to reduce the TeV flux to an unobserved level 
by absorption.
This in turn requires very high luminosities of tertiary photons
in the infrared energy range.
However, the required fluxes are generally orders of magnitude
larger than the observed values (Usov 1994).
We are therefore motivated by the need to contrive an outer gap model
which produces less TeV emission with a moderate infrared luminosity.

High-energy emission from a pulsar magnetosphere,
in fact, crucially depends on the acceleration electric field, 
$\Ell$, along the magnetic field lines.
It was Hirotani \& Shibata (1999a,b; hereafter Paper I, II)
who first solved the spatial distribution of $\Ell$ 
together with particle and $\gamma$-ray distribution functions.
They explicitly demonstrated that 
there is a stationary solution for an outer gap which is formed 
around the null surface at which the 
local Goldreich-Julian charge density 
\begin{equation}
  \rho_{\rm GJ}= \frac{\Omega B_z}{2\pi c [1-(\Omega \varpi/c)^2]}
  \label{def_rhoGJ}
\end{equation}
vanishes,
where $B_z$ is the component of the magnetic field along 
the rotation axis,
$\Omega$ refers to the angular frequency of the neutron star,
$\varpi$ indicates the distance of the point from the rotation axis,
and $c$ is the speed of light.
Subsequently, Hirotani (2000a, hereafter Paper IV) investigated the 
$\gamma$-ray emission from an outer gap,
by imposing a gap closure condition that a single pair
produces one pair in the gap on average.
He demonstrated that $\Ell$ becomes typically less than $10\%$ of the
value assumed in CHR and that the resultant TeV flux is
sufficiently less than the observational upper limit of the pulsed
flux, if the outer gap is immersed in a X-ray field supplied by the 
blackbody radiation from the whole neutron star surface and/or
from the heated polar caps.
In this paper, we develop his method to the case when a magnetospheric
power-law component contributes in addition to the blackbody components.

In the next section, we formulate the gap closure condition.
Solving the condition in \S 3, 
we investigate the acceleration field
and the resultant $\gamma$-ray emissions
as a function of the X-ray field.
In \S 4, we further apply the theory to twelve rotation-powered pulsars 
and predict the absolute fluxes of TeV emission from their outer gaps.
In the final section, we discuss the validity of assumptions
and give some implications on pulse profiles of GeV emissions.

\section{Electrodynamic structure of the gap}

We first describe the acceleration field in \S 2.1,
then consider the energy of curvature-radiated $\gamma$-rays in \S 2.2,
the X-ray field illuminating the gap in \S 2.3
and the pair production mean free path in \S 2.4.
We further formulate in \S 2.5 the gap closure condition 
that one of the copious $\gamma$-rays  emitted by a single pair 
materialize as a pair in the gap on average.
We finally present the resultant $\gamma$-ray properties 
in \S\S 2.6 and 2.7.

\subsection{Acceleration Field in the Gap}

\begin{figure} 
\centerline{ \epsfxsize=9cm \epsfbox[0 60 500 400]{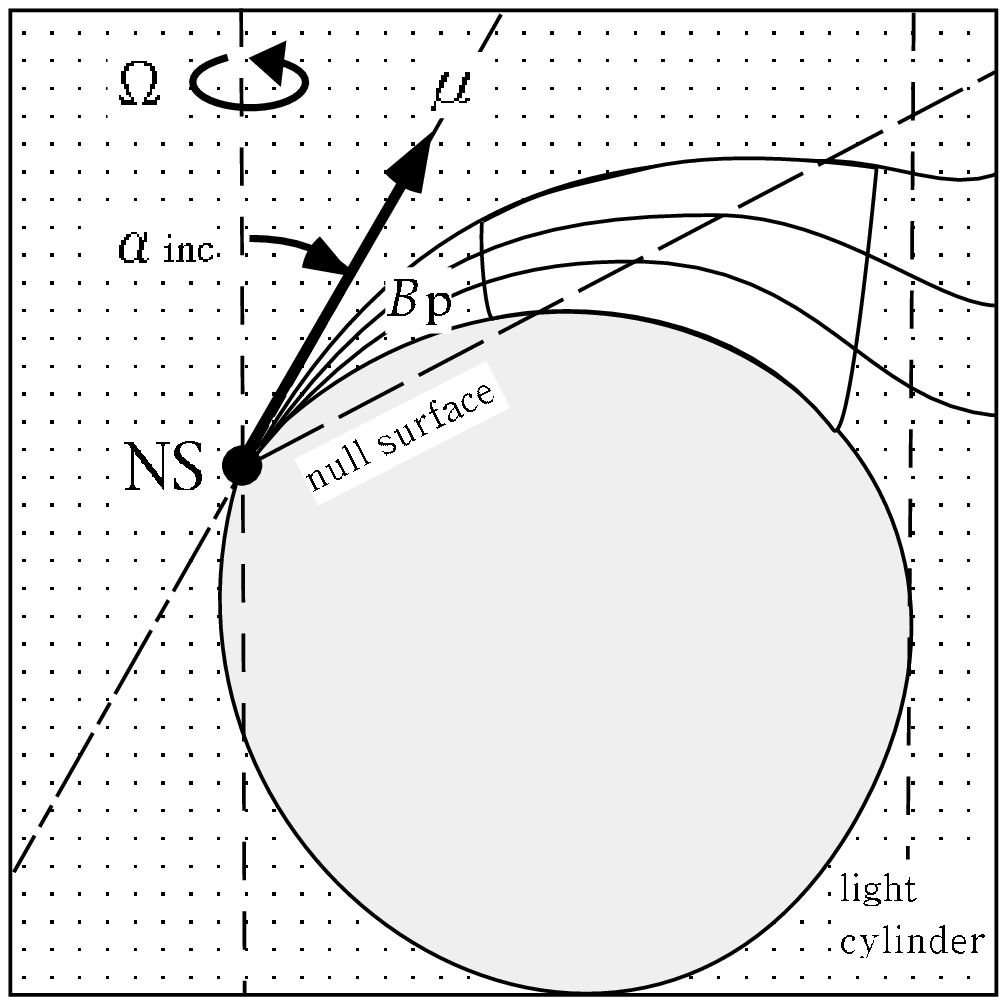} } 
\caption{\label{fig:oblique} 
A side view of a hypothetical outer magnetospheric gap
in which pair production cascade takes place. 
        }
\end{figure}

In this paper, we consider an outer gap in the following 
rectilinear coordinate: 
$x$ is an outwardly increasing coordinate along the magnetic field
lines, while $z$ is parallel to the rotational axis. 
We define $x=0$ to be the intersection between the last open field line
and the null surface where $B_{\rm z}=0$ (fig.~\ref{fig:oblique}).
If we assume that the transfield thickness of the gap is larger than
the gap width along the field lines,
we can neglect $\partial_z{}^2$ term compared with 
$\partial_x{}^2$ one in the Poisson equation
for the non-corotational potential, $\Phi$. 
Then we can Taylor-expand $\rho_{\rm GJ}$ around $x=0$ to obtain 
\begin{equation}
   -\frac{d^2{\Phi}}{dx^2} 
   = - 4\pi A x ,
   \label{eq:Poisson-2}
\end{equation} 
where $A$ is the expansion coefficient of $\rhoGJ$ at $x=0$. 
Since the toroidal current flowing near the light cylinder is unknown,
we simply approximate $B_z$ with its Newtonian value.
Then $A$ is given by
\begin{eqnarray}
  A &\equiv& \frac{3\Omega\mu}{2\pi c r_0{}^4}
             \frac{1}{1-(\Omega r_0 \sin\theta_0/c)^2}
    \nonumber \\
    & & \hspace*{-1.5 truecm}
        \times \left[ \frac32 \sin 2\theta_0 \cos(\theta_0-\inc)
                     +\cos 2\theta_0 \sin(\theta_0-\inc) \right] ,
  \label{eq:def_A}
\end{eqnarray}
where $\mu$ refers to the magnetic dipole moment of the neutron star;
$r_0$ is the distance of the gap center ($x=0$)
from the neutron star, and
$\theta_0$ is the polar angle of the gap center 
(in the first quadrant).
They are related with $\inc$ as 
\begin{equation}
  \frac{r_0}{\rlc} 
  \equiv \frac{\sin^2(\theta_0-\inc)}
              {\sin\theta_{\rm LC}\sin^2(\theta_{\rm LC}-\inc)}
  \label{eq:def-Rnull}
\end{equation}
and
\begin{equation}
  \tan\theta_0
  \equiv \frac{3\tan\inc + \sqrt{9\tan^2\inc+8}}{2},
  \label{eq:def-THnull}
\end{equation}
where the light-cylinder radius is defined by
\begin{equation}
  \rlc = \frac{c}{\Omega} = 3.0 \times 10^8 \Omega_2{}^{-1} \mbox{cm}
\end{equation}
and the colatitude angle $\theta_{\rm LC}$ at which
the last-open fieldline intersects with the light cylinder
is implicitly solved from
\begin{equation}
  2 \cot(\theta_{\rm LC}-\inc)\sin\theta_{\rm LC}+\cos\theta_{\rm LC}=0;
\end{equation}
$\Omega_2 \equiv \Omega / 10^2 \mbox{rad s}^{-1}$.
In the order of magnitude, $A \sim \rhoGJ / \rlc$ holds.
Its exact values are 
$3.19 \times 10^{-12} \Omega_2{}^5 \mu_{30}$ esu 
for $\inc=30^\circ$ and 
$1.22 \times 10^{-11} \Omega_2{}^5 \mu_{30}$ esu 
for $\inc=45^\circ$,
where $\mu_{30} \equiv \mu / 10^{30}\mbox{G cm}^3$. 

Integrating equation (\ref{eq:Poisson-2}),
we obtain the acceleration field 
$\Ell(x) \equiv -d\Phi/dx = \Ell(0) - 2\pi A x^2$,
where $\Ell(0)$ refers to the value of $\Ell$ at $x=0$.
Defining the boundaries of the gap to be the places where
$\Ell$ vanishes, we obtain $\Ell(0)= 2\pi A H^2$.
It is noteworthy that the non-vanishment of 
$d\Ell / dx$ at $x=\pm H$ is consistent with
the stability condition at the plasma-vacuum interface 
if the electrically supported magnetospheric plasma
is completely-charge-separated, 
i.e., if the plasma cloud at $x < -H$ is composed of 
electrons alone (Krause-Polstorff \& Michel 1985a,b).
We assume that the Goldreich-Julian plasma gap boundary 
is stable with $\Ell=0$ on the boundary, $x=-H$.

We can now evaluate the typical strength of $\Ell$ by averaging
its values throughout the gap as follows:
\begin{eqnarray}
  \bar{\Ell} 
    &=& \frac{1}{H} \int_0^H dx \left( \Ell(0)-2\pi A x^2 \right)
  \nonumber \\
    &=& \frac43 \pi A H^2.
  \label{eq:Ell}
\end{eqnarray}
We shall use this $\bar{\Ell}$ as a representative value of the
acceleration field in the gap.
Defining the non-dimensional gap width $h \equiv H/\rlc$ and
substituting the values of $A$, we obtain
\begin{equation}
 \bar{\Ell} = 
   C_{\rm E} \Omega_2{}^3 \mu_{30} h^2 \frac{\rm V}{\rm m},
 \label{eq:Ell_2}
\end{equation}
where $C_{\rm E}=3.61 \times 10^{10}$ for $\inc=30^\circ$, and
$1.39 \times 10^{11}$ for $\inc=45^\circ$.

The voltage drop in the gap is given by
\begin{equation}
  V_{\rm gap} = \bar{\Ell} \cdot 2H
              = \frac83 \pi A H^3.
  \label{eq:Vgap-0}
\end{equation}
For an outer gap which extends to the light cylinder 
($H \sim 0.5 \rlc$),
$V_{\rm gap}$ becomes as large as the available electromotive force
(EMF) exerted on the spinning neutron star surface,
$V_* \approx 10^{15} (\Omega_2/0.5)^2 \mu_{30}$V.

\subsection{Energy of curvature-radiated $\gamma$-rays}

The most effective assumption for the particle motion in the gap
arises from the fact that the velocity saturates immediately 
after their birth in the balance between the radiation reaction force 
and the electric force.
The reaction force is mainly due to curvature radiation if the gap is
immersed in a moderate X-ray field.
Equating the electric force $e\Ell$ and the radiation reaction force,
we obtain the Lorentz factor of saturated particles as follows:
\begin{equation}
  \bar{\Gamma} 
   = \left( \frac{3 R_{\rm c}{}^2 \bar{\Ell}}{2e} 
           \right)^{1/4},
  \label{eq:terminal_0}
\end{equation}
where $R_{\rm c}$ and $e$ refers to the curvature radius of the 
magnetic field line and the magnitude of the charge on the electron.
$\bar{\Gamma}$ is related with $h \equiv H/\rlc$ by
\begin{equation}
 \bar{\Gamma}
   = C_\Gamma R_{0.5}{}^{1/2} (\Omega_2 \mu_{30})^{1/4} h^{1/2}, 
 \label{eq:terminal_2}
\end{equation}
where $C_\Gamma= 9.59 \times 10^7$ for $\inc=30^\circ$, and
$1.34 \times 10^9$ for $\inc=45^\circ$;
$R_{0.5} \equiv R_{\rm c} / (0.5\rlc)$. 

However, if the gap width $H$ is less than the length scale
for the particles to be accelerated up to $\bar{\Gamma}$,
the typical Lorentz factor should be rather estimated by
$e\Ell H/m_{\rm e}c^2$,
because typical particles are accelerated by the potential
$V_{\rm gap}/2=\Ell H$.
Therefore, we can evaluate the Lorentz factor as
\begin{equation}
  \Gamma = \min (\bar{\Gamma}, e\Ell H/m_{\rm e}c^2 )
  \label{eq:terminal}
\end{equation}

Using this $\Gamma$, 
we obtain the central energy of curvature radiation,
\begin{equation}
  E_{\rm c}
     = \frac{3\Gamma^3}{2} \frac{\hbar c}{R_{\rm c}} 
  \label{eq:Ec} 
\end{equation}
where $\hbar$ is the Planck constant divided by $2\pi$.
In this paper,
we adopt the gray approximation in the sense that
all the $\gamma$-rays are radiated at energy $E_{\rm c}$.
In the final section of Paper IV, we demonstrated that
this gray approximation gave a good estimate of the gap width
and other quantities describing the gap,
by comparing them with those obtained in the non-gray cases 
in which the Boltzmann equations of particles and $\gamma$-rays
were solved together with the Poisson equation for $\Phi$.

\subsection{X-ray field}

Before proceeding to the discussion on pair production
mean free path,
it is desirable to describe the X-ray field 
that illuminates the outer gap.
X-ray field of a rotation-powered neutron star 
within the light cylinder
can be attributed to the following three emission processes: \\
(1) Photospheric emission from the whole surface of
a cooling neutron star
(Greenstein \& Hartke 1983; Romani 1987; Shibanov et al. 1992;
 Pavlov et al. 1994; Zavlin et al. 1995). \\
(2) Thermal emission from the neutron star's polar caps 
which are heated by the bombardment of relativistic particles
streaming back to the surface from the magnetosphere
(Kundt \& Schaaf 1993; Zavlin, Shibanov, \& Pavlov 1995;
 Gil \& Krawczyk 1996). \\
(3) Non-thermal emission from relativistic particles accelerated 
in the pulsar magnetosphere
(Ochelkov \& Usov 1980a,b; El-Gowhari \& Arponen 1972;
 Aschenbach \& Brinkmann 1974; Hardee \& Rose 1974; Daishido 1975).

The spectrum of the first component is expected to be 
expressed with a modified blackbody.
However, for simplicity, we approximate it in terms of a Plank function 
with temperature $kT_{\rm s}$, 
because the X-ray spectrum is occasionally fitted 
by a simple blackbody spectrum.
We regard a blackbody component as the first one
if its observed emitting area is
comparable with the whole surface of a neutron star,
$A_* \equiv 4 \pi r_*{}^2$,
where $r_*$ denotes the neutron star radius. 
We take account of both the pulsed and the non-pulsed 
surface blackbody emission 
as this soft blackbody component.

As for the second component, we regard a blackbody component as
the heated polar cap emission if its observed emitting area
is much smaller than $A_*$.
We approximate its spectrum by a Planck function.
We take account of both the pulsed and the non-pulsed 
polar cap emission 
as this hard blackbody component.

Unlike the first and the second components, 
a power-law component is usually dominated by a nebula emission.
To get rid of the nebula emission, 
which illuminates the outer gap inefficiently,
we adopt only the pulsed components of a power-law emission
as the third component.

\subsection{Pair Production Mean Free Path}

In this subsection,
we draw attention to how the pair production 
mean free path is related with the X-ray field described in the 
previous section.
To this end, we first give the threshold energy for X-ray photons
to materialize.
Then, we consider the mean free path
for a $\gamma$-ray photon to materialize as a pair in a 
collision with one of the soft blackbody X-rays in \S~\ref{sec:softBB},
the hard blackbody ones in \S~\ref{sec:hardBB},
and the magnetospheric, power-law X-rays in \S~\ref{sec:PW}.
We finally summarize how the real mean free path 
should be computed under the existence of these three X-ray
components in \S~\ref{sec:mfp}.

\subsubsection{Threshold Energy}
\label{sec:Eth}

The threshold energy for a soft photon to materialize as
a pair in a collision with the $\gamma$-ray having energy
$m_{\rm e}c^2 \epsilon_\gamma = E_{\rm c}$, is given by
\begin{equation}
  E_{\rm th} 
     = \frac{2}{1-\cos\phi_{\rm ab}\cos\theta_{\rm c}}
       \frac{m_{\rm e}c^2}{\Eg},
  \label{eq:Eth} 
\end{equation}
where $\cos\phi_{\rm ab}\cos\theta_{\rm c}$ refers to 
the cosine of the three-dimensional 
collisional angle between the X-ray and the $\gamma$-ray.
The lower bould of 
$2/(1-\cos\phi_{\rm ab}\cos\theta_{\rm c})$ is unity, 
which is realized if the two photons head-on collide.

To evaluate $\cos\phi_{\rm ab}\cos\theta_{\rm c}$, 
we must consider $\gamma$-ray's toroidal  momenta due to aberration.
At the gap center,
the aberration angle $\phi_{\rm ab}$ is given by
$\sin^{-1}(r_0 \sin\theta_0 / \rlc)$.
In the case of $\inc=30^\circ$,
we obtain $\phi_{\rm ab}=20.4^\circ$.
The collisional angle on the poloidal plane becomes
$\theta_{\rm c}=90^\circ-\theta_0 = 21.6^\circ$
(or $\theta_{\rm c}=90^\circ+\theta_0 = 158.4^\circ$)
for outwardly (or inwardly) propagating $\gamma$-rays,
where $\theta_0$ is computed from equation (\ref{eq:def-THnull}).
Therefore, we obtain
$\cos\phi_{\rm ab} \cos\theta_{\rm c}= \pm 0.855$,
where the upper and the lower sign correspond to the 
outwardly and inwardly propagating $\gamma$-rays, respectively.
On the other hand, in the case of $\inc=45^\circ$, 
we have $\phi_{\rm ab}=14.0^\circ$ and $\theta_{\rm c}= 15.6^\circ$ 
(or $\theta_{\rm c}= 164.3^\circ$)
to obtain $\cos\phi_{\rm ab}\cos\theta_{\rm c}= \pm 0.923$.

\subsubsection{Soft Blackbody Component}
\label{sec:softBB}

We first consider 
the pair production mean free path, $\lambda_{\rm s}$,
for a $\gamma$-ray photon to materialize in a collision with 
one of the {\it soft} blackbody X-rays.
In a realistic outer gap, both the outwardly and inwardly
propagating $\gamma$-rays contribute for $\lambda_{\rm s}$.
Therefore, $\lambda_{\rm s}$ is given by
an arithmetic average of these two contributions as follows
(Paper IV; see also eq. [3.1] in Blandford \& Levinson 1995
 for the factor $1\pm\mu_{\rm c}$): 
\begin{eqnarray}
  \frac{1}{\lambda_{\rm s}}
  &\equiv& 
  \kappa (1-\mu_{\rm c})
     \int_{2/(1-\mu_{\rm c})\Eg}^\infty d\epsilon_{\rm x}
     \frac{dN_{\rm s}}{d\Ex} \sgP(\Eg, \Ex,\mu_{\rm c})
  \nonumber \\
  & & \hspace*{-1.5 truecm} +
  (1-\kappa) (1+\mu_{\rm c})
     \int_{2/(1+\mu_{\rm c})\Eg}^\infty d\epsilon_{\rm x}
     \frac{dN_{\rm s}}{d\Ex} \sgP(\Eg, \Ex,-\mu_{\rm c}),
  \nonumber \\
  & & \ 
  \label{eq:def_lambda_4}
\end{eqnarray}
where 
$\mu_{\rm c} \equiv \vert \cos\phi_{\rm ab}\cos\theta_{\rm c} \vert$; 
the pair production cross section is given by
(Berestetskii et al. 1989)
\begin{eqnarray}
  & & \sgP(\Eg,\Ex,\mu_{\rm c})
  \nonumber \\
  &\equiv& \frac{3}{16} \sgT
    (1-v^{2}) \left[ (3-v^4) \ln \frac{1+v}{1-v} -2v(2-v^{2}) \right],
  \nonumber
  \label{eq:def-sgP}
\end{eqnarray}
\begin{equation}   
  v(\Eg, \Ex, \mu_{\rm c}) 
     \equiv \sqrt{ 1 - \frac{2}{1-\mu_{\rm c}} \frac{1}{\Eg\Ex} },
  \label{eq:def-v}
\end{equation} 
where $\sgT$ is the Thomson cross section,
and $\Ex \equiv E_{\rm x}/m_{\rm e}c^2$
refers to the dimensionless energy of an X-ray photon.
We may notice here that the non-dimensional threshold energy 
($E_{\rm th}/m_{\rm e}c^2$) appears in the
lower bound of the integral in equation (\ref{eq:def_lambda_4}).

When a blackbody emission dominates the X-ray field,
$H$ occasionally becomes as large as $0.3 \rlc$.
In this case, we must take the dilution effect of the
surface radiation into account.
The number density of the soft X-rays between energies
$m_{\rm e}c^2 \Ex$ and $m_{\rm e}c^2 (\Ex+d\Ex)$
at position $x$ is given by
\begin{equation}
  \frac{dN_{\rm s}}{d\epsilon_{\rm x}}
  =  \left[ 1 +2\sin\theta_0 \frac{x}{\rlc}
              +\left( \frac{x}{\rlc} \right)^2 \right]^{-1}
     \left( \frac{dN_{\rm s}}{d\epsilon_{\rm x}} \right)_0 .
  \label{eq:def_Ns}
\end{equation}
Here, the number density at the gap center is given by the Planck law
\begin{equation}
  \left(\frac{dN_{\rm s}}{d\epsilon_{\rm x}}\right)_0
  = \frac{1}{4\pi^2} 
           \left( \frac{m_{\rm e}c^2}{c \hbar} \right)^3
           \left( \frac{A_{\rm s}}{4\pi r_0^2} \right)
      \frac{\Ex{}^2}
           {\exp(\Ex/\Delta_{\rm s})-1},
  \label{eq:def_Ns0}
\end{equation}
where
$A_{\rm s}$ indicates the observed emission area of the soft blackbody;
$\Delta_{\rm s}$ is defined by
\begin{equation}
  \Delta_{\rm s} \equiv \frac{k T_{\rm s}}{m_{\rm e}c^2},
  \label{eq:def_Ds}
\end{equation}
where $kT_{\rm s}$ refers to the 
soft blackbody temperature measured by a distant observer.
Since the outer gap is located outside of the deep gravitational
potential well of the neutron star, 
the photon energy there is essentially the same as the distant observer
measures.

The first (or the second) term in equation (\ref{eq:def_lambda_4})
represents the contribution from the
outwardly (or inwardly) propagating $\gamma$-rays;
For $\inc=30^\circ$, 
we adopt  $\mu_{\rm c}=0.855$,
whereas for $45^\circ$, $\mu_{\rm c}=0.923$.
The weight $\kappa$ reflects the ratio of the fluxes between the 
outwardly and inwardly propagating $\gamma$-rays.
For example, if $\kappa$ were to be $1.0$,
only outwardly propagating $\gamma$-rays would contribute for the
pair production.
From Paper III, we know that the flux of the outwardly propagating 
$\gamma$-rays is typically about ten times larger than the
inwardly propagating ones.
In what follows, we thus adopt $\kappa=0.9$.
This value of $\kappa$ was justified in the final section of Paper IV. 

\subsubsection{Hard Blackbody Component}
\label{sec:hardBB}

In this subsection, 
let us consider the case when the X-ray field is dominated by
the {\it hard} blackbody component.
In the same manner as the soft blackbody component, 
the {\it hard} blackbody component gives the following 
mean free path for pair production:
\begin{eqnarray}
  \frac{1}{\lambda_{\rm h}}
  &\equiv& 
  \kappa (1-\mu_{\rm c})
     \int_{2/(1-\mu_{\rm c})\Eg}^\infty d\epsilon_{\rm x}
     \frac{dN_{\rm h}}{d\Ex} \sgP(\Eg, \Ex,\mu_{\rm c})
  \nonumber \\
  & & \hspace*{-1.5 truecm} +
  (1-\kappa) (1+\mu_{\rm c})
     \int_{2/(1+\mu_{\rm c})\Eg}^\infty d\epsilon_{\rm x}
     \frac{dN_{\rm h}}{d\Ex} \sgP(\Eg, \Ex,-\mu_{\rm c}),
  \nonumber \\
  & & \ 
  \label{eq:def_lambda_5}
\end{eqnarray}
where $\mu_{\rm c}$ takes the same value as the 
soft-blackbody-dominated case
and $\kappa=0.9$.
The number density of the hard blackbody X-rays between energies
$m_{\rm e}c^2 \Ex$ and $m_{\rm e}c^2 (\Ex+d\Ex)$
at position $x$ is given by 
\begin{equation}
  \frac{dN_{\rm h}}{d\epsilon_{\rm x}}
  =  \left[ 1 +2\sin\theta_0 \frac{x}{\rlc}
              +\left( \frac{x}{\rlc} \right)^2 \right]^{-1}
     \left( \frac{dN_{\rm h}}{d\epsilon_{\rm x}} \right)_0 ,
  \label{eq:def_Nh}
\end{equation}
\begin{equation}
  \left( \frac{dN_{\rm h}}{d\epsilon_{\rm x}} \right)_0
  = \frac{1}{4\pi^2} 
           \left( \frac{m_{\rm e}c^2}{c \hbar} \right)^3
           \left( \frac{A_{\rm h}}{4\pi r_0^2} \right)
      \frac{\Ex{}^2}
           {\exp(\Ex/\Delta_{\rm h})-1},
  \label{eq:def_Nh0}
\end{equation}
where $A_{\rm h}$ denotes the observed emission area 
of the  hard-blackbody emission.
$\Delta_{\rm h}$ is defined by
\begin{equation}
  \Delta_{\rm h} \equiv \frac{k T_{\rm h}}{m_{\rm e}c^2},
\end{equation}
where $kT_{\rm h}$ refers to the 
hard blackbody temperature measured by a distant observer.

\subsubsection{Power-law Component}
\label{sec:PW}

Since the secondary photons emitted 
outside of the gap via synchrotron process 
will be beamed in the same direction of the primary $\gamma$-rays,
the typical collision angle on the poloidal plane
can be approximated as $h \equiv H/\rlc$ rad.
It follows that the mean free path corresponding to the
power-law emission becomes
\begin{equation}
  \frac{1}{\lambda_{\rm pl}}
  = (1-\mu_{\rm c})
    \int_{\zeta/\epsilon_\gamma}^\infty d\epsilon_{\rm x}
      \frac{dN_{\rm pl}}{d\epsilon_{\rm x}}
      \sgP(\epsilon_\gamma, \epsilon_{\rm x},\mu_{\rm c}),
  \label{eq:def_lambda_3}
\end{equation}
where 
\begin{equation}
  \zeta \equiv \frac{2}{1-\mu_{\rm c}};
\end{equation}
$dN_{\rm pl}/d\Ex$ refers to the number density of the power-law X-rays
between energies $m_{\rm e}c^2 \Eg$ and $m_{\rm e}c^2 (\Eg +d\Eg)$; 
we adopt
\begin{equation}
  \mu_{\rm c}= \cos \left( \frac{H}{r_0\sin\theta_0} \phi_{\rm ab} \right) 
               \cos h
\end{equation}
for the power-law component.
We may notice here that both the $\gamma$-rays and the magnetospheric,
power-law X-rays suffer aberration
and that the resultanlt collision angle in the azimuthal direction
is less than $\phi_{\rm ab}$ and can be assumed to be
$(H/r_0\sin\theta_0)\phi_{\rm ab}$.
The dependence of $dN_{\rm pl}/d\Ex$ on $x$ is unknown.
We thus simply assume that it is constant throughout the gap 
and specify the form as
\begin{equation}
  \frac{dN_{\rm pl}}{d\epsilon_{\rm x}}
    = N_{\rm pl} \epsilon_{\rm x}{}^\alpha
  \quad (\epsilon_{\rm min} < \epsilon_{\rm x} < \epsilon_{\rm max}).
  \label{eq:def-Npl}
\end{equation}
The photon index $\alpha$ is typically between $-2$ and $-1$
for a pulsed, power-law X-ray component in hard X-ray band
(e.g., Saito 1998).

\subsubsection{Pair Production Mean Free Path}
\label{sec:mfp}

The true mean path, $\lambda_{\rm p}$,
is determined by the component that dominates the X-ray field,
or equivalently, the smallest one among 
$\lambda_{\rm s}$, $\lambda_{\rm h}$, and $\lambda_{\rm pl}$.
Therefore, we can reasonably evaluate $\lambda_{\rm p}$ as 
\begin{equation}
  \frac{1}{\lambda_{\rm p}}
  = \frac{1}{\lambda_{\rm s}}
   +\frac{1}{\lambda_{\rm h}}
   +\frac{1}{\lambda_{\rm pl}}.
    \label{eq:mfp}
\end{equation}

When a pulsar is young, the third term will dominate because of its
strong magnetospheric emission.
As the pulsar evolves, 
the first term will become dominant owing to the diminishing magnetospheric
emission.
As the pulsar evolves further,
polar cap heating due to particle bombardment begins to dominate;
therefore, the second term becomes important.

\subsection{Gap Closure}

The gap width $2H$ is adjusted
so that a single pair produces copious 
$\gamma$-ray photons (of number $N_\gamma$) 
one of which materializes as a pair on average.
Since a typical $\gamma$-ray photon runs the length $H$ in the gap
before escaping from either of the boundaries,
the probability of a $\gamma$-ray photon to 
materialize within the gap, $N_\gamma{}^{-1}$, 
must coincide with the optical depth for absorption, 
$H/\lambda_{\rm p}$. 
Considering the position dependence of $\lambda_{\rm p}$ on $x$,
we obtain 
\begin{equation}
  \frac{1}{N_\gamma} =
  \frac{1}{2} \int_{-H}^{H} \frac{dx}{\lambda_{\rm p}},
  \label{eq:closure_0}
\end{equation}
where 
\begin{equation}
  N_\gamma 
    \approx \frac{H}{c} \frac{4 e^2\Gamma}{9 \hbar R_{\rm c}}.
  \label{eq:Ngamma}
\end{equation}
Here, $\lambda_{\rm p}$ is given by equation (\ref{eq:mfp}).
Equation (\ref{eq:Ngamma}) is derived as follows:
A single $e^+$ or $e^-$ emits $\gamma$-rays at the rate 
$P= 2ce^2 \Gamma^4/(3R_{\rm c}{}^2)$.
Dividing $P$ with the averaged photon energy,
$3\Gamma^3 \hbar c/(2 R_{\rm c})$,
we can estimate the number of $\gamma$-rays emitted per unit time
by a single $e^+$ or $e^-$, $\dot{N}_\gamma$.
Noting that a typical particle runs length $H$ in the gap on average,
we obtain $N_\gamma=(H/c)\dot{N}_\gamma$,
which reduces to equation (\ref{eq:Ngamma}).

Substituting equations (\ref{eq:def_Ns}), (\ref{eq:def_Nh}), and
(\ref{eq:mfp}) into (\ref{eq:closure_0}), we obtain
the gap closure condition
\begin{equation}
  \frac{1}{N_\gamma}
  = \left( \frac{H}{\lambda_{\rm s,0}}
          +\frac{H}{\lambda_{\rm h,0}}
    \right) I(h)
   +\frac{H}{\lambda_{\rm pl}},
  \label{eq:closure}
\end{equation}
where $\lambda_{\rm s,0}$ and $\lambda_{\rm h,0}$ refer to the
values of $\lambda_{\rm s}$ and $\lambda_{\rm h}$ at $x=0$, and 
\begin{equation}
  I(h) \equiv
    \frac{ \tan^{-1}(\tan\theta_0+h\sec\theta_0)
          -\tan^{-1}(\tan\theta_0-h\sec\theta_0) }
         {2h\cos\theta_0}.
  \label{eq:def_I}
\end{equation}
For a very thin gap ($h=H/\rlc \ll 1$), $I(h)$ approaches to unity.
When the surface blackbody components dominate the X-ray field,
$h$ can be as large as $0.3$;
as a result, the $r^{-2}$ variation of the X-ray density 
becomes important.
We take account of this effect in function $I(h)$.
When the power-law component dominates, on the other hand,
$h \ll 1$ is usually satisfied 
(see discussion in \S~\ref{sec:gamma-ray});
therefore, we only evaluate the X-ray density at the gap center
in equation~(\ref{eq:closure}).
Combining equations (\ref{eq:Ngamma}) and (\ref{eq:closure}), 
and representing $\Gamma$, $\Eg=E_{\rm c}/m_{\rm e}c^2$ 
in terms of $h \equiv H/\rlc$,
we finally obtain the equation describing
$h$ as a function of 
$B_5$, $\Omega_2$, $kT_{\rm s}$, $A_{\rm s}$,
$kT_{\rm h}$, and $A_{\rm h}$.
Once $h$ is solved, other quantities such as $\bar{\Ell}$, $\Gamma$,
$E_{\rm c}$ can be computed straightforwardly.

\subsection{Luminosity of GeV emissions}
\label{sec:Lgev}

Let us first consider the luminosities of curvature-radiated
$\gamma$-rays.
The luminosity, $L_{\rm GeV}$, can be estimated by multiplying 
the total number of positrons and electrons in the gap, $N_{\rm e}$,
the number of $\gamma$-rays emitted per particle per unit time, 
$N_\gamma/(H/c)$,
and $\gamma$-ray energy, $E_{\rm c}$.
Supposing the conserved current density is 
$\chi c \rhoGJ = \chi (\Omega B/2\pi e)$,
and assuming that the gap exists $\pi (H/\rlc)$ rad
in the azimuthal direction,
we obtain
\begin{equation}
  N_{\rm e}\sim \chi \frac{\Omega B}{2\pi ce}
                \cdot \frac{\pi H}{\rlc}
                \cdot r_0 \sin\theta_0 \cdot D_\perp \cdot 2H.
  \label{eq:Ne}
\end{equation}
where $D_\perp$ is the transfield thickness of the gap on the poloidal
plane.
The distance of the center of the gap from the rotation axis,
$r_0 \sin\theta_0$, becomes $0.393\rlc$ for $\inc=30^\circ$,
and $0.285 \rlc$ for $\inc=45^\circ$.
The strength of magnetic field at the gap center
can be given by 
\begin{eqnarray}
  B &=& 3.70 \times 10^5 \sqrt{1+3\cos^2(\theta_0-\inc)}
          \left( \frac{r_0}{\rlc} \right)^{-3} 
    \nonumber \\
    && \qquad \Omega_2{}^3 \mu_{30}
       \qquad \mbox{G}.
  \label{eq:B}
\end{eqnarray}
As discussed in \S 2.4 in Paper IV,
we adopt $\chi=0.1$ and $D_\perp=0.3 \rlc$ as typical values
in this paper.

From equations (\ref{eq:Ngamma}) and (\ref{eq:Ec}),  
we obtain
\begin{equation}
  L_{\rm GeV}
    = \frac23 \left(\chi\frac{D_\perp}{\rlc}\right)
      \frac{e\Omega B r_0 \sin\theta_0}
           {(R_{\rm c}/\rlc)^2}
      \Gamma^4 h^2,
  \label{eq:L_GeV_1}
\end{equation}
where $\Gamma$ is given by equation (\ref{eq:terminal}).
If $\Gamma$ is saturated, equation (\ref{eq:terminal_0}) or
(\ref{eq:terminal_2}) gives
\begin{equation}
  L_{\rm GeV}
    \sim C_{\rm GeV} \, \left(\chi \frac{D_\perp}{\rlc} \right)
             \Omega_2{}^4 \mu_{30}{}^2 h^4
    \mbox{ergs s}^{-1}.
    \label{eq:L_GeV_2}
\end{equation}
where $C_{\rm GeV}=1.19 \times 10^{39}$ for $\inc=30^\circ$, and
$1.20 \times 10^{40}$ for $\inc=45^\circ$.

\subsection{Luminosity of TeV emissions}
\label{sec:Ltev}

The relativistic particles produce $\gamma$-rays mainly via
curvature radiation as described in the preceding sections.
However, even though energetically negligible,
it is useful to draw attention to the TeV $\gamma$-rays produced
via inverse-Compton (IC) scatterings.
Since the particles' Lorentz factor becomes $\sim 10^{7.5}$,
it is the infrared photons with energy 
$\sim 0.01$ eV that contribute most effectively 
as the target photons of IC scatterings.
Neither the higher energy photons like surface blackbody X-rays 
nor the lower energy photons like polar-cap radio emission
contribute as the target photons,
because they have either too small cross sections 
or too small energy transfer when they are scattered. 
On these grounds, we obtain the following upper bound for the
luminosity of the IC-scattered $\gamma$-rays:
\begin{eqnarray}
  L_{\rm TeV}
    &<& 5.63 \times 10^{25} B_5 L_{30} 
             \left( \frac{\rlc}{r_0}\sin\theta_0 \right)
             \left( \chi \frac{D_\perp}{\rlc} \right)
  \nonumber \\
  & & \hspace*{3.0 truecm}
      \Gamma h^2 \mbox{ergs s}^{-1},
  \label{eq:L_TeV_1}
\end{eqnarray}
where $L_{30}$ refers to the luminosity of infrared photons
that can be scattered up to TeV energy range in the unit of
$10^{30} \mbox{ergs s}^{-1}$;
$B_5 \equiv B/(10^5 {\rm G})$.
The inequality comes from the fact that the scattered $\gamma$-rays
cannot have energies greater than $\Gamma m_{\rm e} c^2$.

The $\nu F_\nu$ flux of IC-scattered $\gamma$-rays, 
$(\nu F_\nu)_{\rm TeV}$ [Jy~Hz],
can be readily computed as
\begin{equation}
  (\nu F_\nu)_{\rm TeV} =
    10^{-20} \left( \frac{L_{\rm TeV}}{\mbox{ergs s}^{-1}} \right)
             \left[ \frac{\omega_{\rm TeV}}{1 \,\mbox{ster}}
             \right]^{-1}
             d_1{}^{-2}
    \mbox{Jy Hz},
  \label{eq:TeV_flux_1}
\end{equation}
where $\omega_{\rm TeV}$ refers to the solid angle 
in which the TeV $\gamma$-rays are emitted,
and $d_1 \equiv d / 1 \mbox{kpc}$.

An analogous relation holds for an infrared flux, 
$(\nu F_\nu)_{0.01{\rm eV}}$,
if the infrared luminosity, $L_{0.01{\rm eV}}$, 
is emitted in a solid angle $\omega_{0.01{\rm eV}}$ ster.
We thus obtain the flux ratio
\begin{eqnarray}
  \frac{(\nu F_\nu)_{\rm TeV}}{(\nu F_\nu)_{0.01{\rm eV}}}
  &=& 5.63 \times 10^{-5} 
      \left( \frac{\rlc}{r_0} \sin\theta_0 \right)  
  \nonumber \\
  & & \hspace*{-1.0 truecm}
      \times \Gamma \left( \chi \frac{D_\perp}{\rlc} \right)
             \frac{B}{10^5\mbox{G}}
             \frac{\omega_{0.01{\rm eV}}}{\omega_{\rm TeV}} h^2.
  \label{eq:TeV_flux_2}
\end{eqnarray}
Since we take the ratio of $\nu F_\nu$ flux at two different
energies, the uncertainties arising from the distance disappears
on the right-hand side.
It follows from equation (\ref{eq:TeV_flux_2}) that
the ratio becomes at most $10^3$ in order of magnitude,
because $\chi(D_\perp/\rlc) < 1$, $h<0.5$, and $\Gamma \sim 10^{7.5}$
hold in general.

\section{X-ray field vs. $\gamma$-ray emission}
\label{sec:X-gamma}

Before proceeding to an application to individual pulsars,
it will be useful to investigate the general properties of 
$\gamma$-ray emission as a function of the X-ray field illuminating 
the gap.
To this aim, we first demonstrate how the gap width $h \equiv H/\rlc$
depends on the X-ray field in \S 3.1,
by solving the gap closure condition (\ref{eq:closure}).
We then present the energies and the luminosities of 
the curvature-radiated and the IC-scattered $\gamma$-rays in \S 3.2.
Throughout this section, we adopt 
$\Omega_2=0.5$, $\mu_{30}=3.0$, 
$A_{\rm s}=A_* \equiv 4\pi r_*{}^2$, $kT_{\rm h}=200$eV, $\alpha=-1.5$,
$\epsilon_{\rm min}= 0.1 \mbox{keV} / 511 \mbox{keV} = 0.00020$, 
and $\epsilon_{\rm max}= 100 \mbox{keV} / 511 \mbox{keV} = 0.20$.

\subsection{Gap width}
\label{sec:width}

\begin{figure} 
\centerline{ \epsfxsize=8.5cm \epsfbox[200 2 700 350]{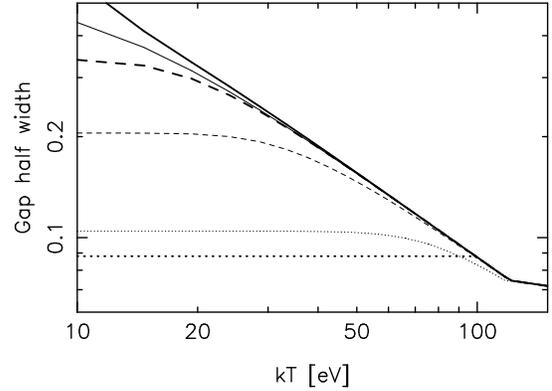} } 
\caption{\label{fig:width} 
Examples of the gap half width
divided by light cylinder radius
as a function of $kT_{\rm s}$ [eV].
The input parameters of the X-ray field of each curve are 
listed in table~1.
For all the six curves, pulsar parameters are fixed as
$\Omega=50 \mbox{ rad s}^{-1}$, $\mu=10^{30} \mbox{G cm}^3$,
and $\inc=30^\circ$.
        }
\end{figure}

The results of the gap half width divided by the light cylinder radius
are presented in figure~\ref{fig:width}.
The abscissa is $kT_{\rm s}$ in eV.
The model parameters of the X-ray field for the six curves
are summarized in table~1.
For the three thick curves, the hard blackbody component is not 
included (i.e., $A_{\rm h}=0$).
The thick solid, dashed, and dotted curves correspond to 
$N_{\rm pl}=0, 10^{10}, 10^{16}$, respectively.
Therefore, the thick solid curve corresponds to the least dense
X-ray field thereby gives the greatest $h \equiv H/\rlc$
for a specific value of $kT_{\rm s}$.
For the three thin curves, on the other hand, 
$N_{\rm pl}$ is set to be $0$,
so that the power-law component does not contribute.
The thin solid, dashed, and dotted curves correspond to
$A_{\rm h} / A_*= 10^{-4}, 10^{-3}, 10^{-2}$, respectively.

\noindent
\begin{table*}
  \centering
    \begin{minipage}{140mm}
      \caption{Model parameters of X-ray field}
      \begin{tabular}{@{}lccccccc@{}}
        \hline
        \hline
        \	& $A_{\rm s}/A_*$
		& $kT_{\rm h}$
		& $A_{\rm h}/A_*$
		& $N_{\rm pl}$
		& $-\alpha$
		& $E_{\rm min}$
		& $E_{\rm max}$		\\
        \	& 
		& eV
		& 
		& cm${}^{-3}$
		& 
		& keV
		& keV			\\
        \hline
        thick solid
		& 1.0
		& 200
		& 0
		& 0
		& 2.0
		& 0.1
		& 100			\\
        thick dashed
		& 1.0
		& 200
		& 0
		& $10^{10}$
		& 2.0
		& 0.1
		& 100			\\
        thick dotted
		& 1.0
		& 200
		& 0
		& $10^{16}$
		& 2.0
		& 0.1
		& 100			\\
        thin solid
		& 1.0
		& 200
		& $10^{-4}$
		& 0
		& 2.0
		& 0.1
		& 100			\\
        thin dashed
		& 1.0
		& 200
		& $10^{-3}$
		& 0
		& 2.0
		& 0.1
		& 100			\\
        thin dotted
		& 1.0
		& 200
		& $10^{-2}$
		& 0
		& 2.0
		& 0.1
		& 100			\\
        \hline
      \end{tabular}
    \end{minipage}
\end{table*}

First of all, it follows from the figure that $H$ 
(or equivalently $h$ for a fixed $\Omega$) 
is a decreasing function of $kT_{\rm s}$.
The reason is as follows: If $kT_{\rm s}$ increases, 
the number density of target soft photons above threshold energy
$N_{\rm s}(E>E_{\rm th})$ increases 
for a fixed value of $E_{\rm th}$.
The increased $N_{\rm s}(E>E_{\rm th})$ results in 
the decrease of $\lambda_{\rm p}$,
which reduces $H$ (eq. [\ref{eq:closure}]).
Accurately speaking, 
the reduced $H$ results in a decrease of $\Ell$ and hence $E_{\rm c}$,
thereby increases $E_{\rm th}$. 
The increased $E_{\rm th}$ decreases
$N_{\rm s}(E>E_{\rm th})$ to partially cancel the initial decrease of
$\lambda_{\rm p}$.
In addition, the reduction of $H$ implies the reduction of
the emitting length for a particle, thereby decreases $N_\gamma$
and partially cancel the initial decrease of $H$
(see eq. [\ref{eq:closure}]).
Nevertheless, both of the two effects are passive;
therefore, the nature of the decrease of $H$ with increasing 
$kT_{\rm s}$ is unchanged.

The second thing to note is that 
$h$ decreases with increasing $A_{\rm h}$, 
as indicated by the three thin curves in figure \ref{fig:width}.
This is because when $A_{\rm h}$ increases
the number density of the hard blackbody component,
$N_{\rm h}(E_{\rm x}>E_{\rm th})$, increases.
This in turn leads to the decrease of $\lambda_{\rm p}$,
which results in the decrease of $h$.
If $A_{\rm H}$ is as small as $10^{-4} A_*$
(the thin solid line),
the X-ray field is dominated by the power-law component
only in the small $kT_{\rm s}$ range;
this can be understood because the thin solid line
significantly deviates from the thick solid line at $kT_{\rm s}<30$eV.
However, if $A_{\rm h}$ is as large as $10^{-2}$
(the thin dotted line),
the X-ray field is dominated by the hard blackbody component
up to very high $kT_{\rm s}$ ($ \sim 100$eV).

The third thing is that $h$ decreases with
increasing $N_{\rm pl}$.
For a very strong magnetospheric emission (the thick dotted line),
$h$ becomes not more than $0.1$.

In short, the gap width is a decreasing function of
of the X-ray number density,
regardless of the component that dominates the X-ray field.

\subsection{Gamma-ray Luminosities}
\label{sec:gamma-lum}

\begin{figure} 
\centerline{ \epsfxsize=8.5cm \epsfbox[200 2 700 350]{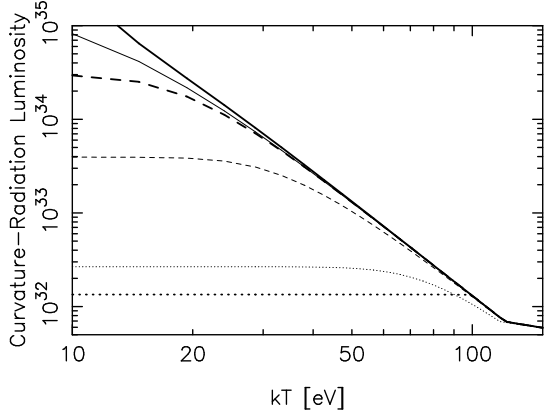} } 
\caption{\label{fig:Lgev} 
Luminosity [ergs s${}^{-1}$] of the curvature-radiated $\gamma$-rays.
The abscissa represents $kT_{\rm s}$ in eV unit.
The six curves correspond to the same parameter set as in figure~2
(or table~1).
        }
\end{figure}

Let us now consider the luminosities of 
the curvature-radiated $\gamma$-rays.
Substituting the results of $h$ into (\ref{eq:L_GeV_1}), 
we obtain $L_{\rm GeV}$; in figure~\ref{fig:Lgev},
we present the results for $\chi=0.1$ and $D_\perp=0.3\rlc$.
It follows from this figure that 
$L_{\rm GeV}$ is a decreasing function of 
the X-ray number density, regardless of 
the component that dominates the X-ray field.
This is because the increase of the target X-ray photons
results in the decreases of $H$.

\begin{figure} 
\centerline{ \epsfxsize=8.5cm \epsfbox[200 2 700 350]{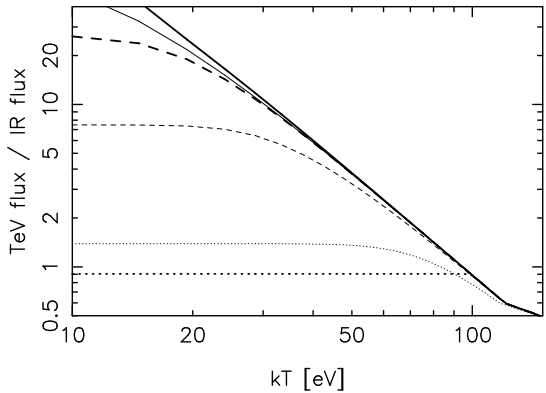} }
\caption{\label{fig:Ltev} 
Ratio of the fluxes between TeV and infrared energy ranges.
The abscissa represents $kT_{\rm s}$ in eV unit.
The TeV $\gamma$-rays are radiated by the primary positrons and
electrons in the gap via IC scatterings,
while the infrared photons are presumably emitted by
tertiary pairs outside of the gap via synchrotron process.
The six curves correspond to the same parameter set as in figure~2
(or table~1).
        }
\end{figure}

We next present the expected ratio of the TeV and the infrared fluxes
in figure~\ref{fig:Ltev}, by using equation (\ref{eq:TeV_flux_2}). 
It follows that the $\nu F_\nu$ flux of TeV emission will not exceed 
$10^{11} \mbox{Jy Hz}$ for a typical infrared flux 
($< 10^9 \mbox{Jy Hz}$).
Therefore, we can conclude that the pulsed TeV emission 
from rotation-powered pulsars
cannot be detected in general by the current ground-based telescopes.

\section{Application to Individual Pulsars} 
\label{sec:appl}

In this section, we apply the theory to the eleven rotation-powered
pulsars of which X-ray field at the outer gap can be deduced 
from observations.
We first describe their X-ray and infrared fields 
in the next two subsections,
and present resultant GeV and TeV emissions from individual pulsars
in \S~4.3 and 4.4.

\subsection{Input X-ray Field}
\label{sec:input_Xray}

We present the observed X-ray properties of individual pulsars
in order of spin-down luminosity, $\dot{E}_{\rm rot}$ (table~2).
We assume 
$\epsilon_{\rm min}=0.1 \mbox{keV} / 511 \mbox{keV}$ and
$\epsilon_{\rm max}=100 \mbox{keV} / 511 \mbox{keV}$
for homogeneous discussion.

\begin{table*}
  \centering
    \begin{minipage}{140mm}
      \caption{Input X-ray field}
      \begin{tabular}{@{}lccccccccc@{}}
        \hline
        \hline
        pulsar	
		& distance
		& $\Omega$		& $\log_{10}\mu$
		& $kT_{\rm s}$		& $A_{\rm s}/A_*$      
		& $kT_{\rm h}$		& $A_{\rm h}/A_*$      
		& $N_{\rm pl}$		& $-\alpha$		\\
        \	
		& kpc
		& rad s${}^{-1}$	& lg(G cm${}^3$)
		& eV			&
		& eV			& 
		& cm${}^{-3}$		&			\\
        \hline
        Crab	
		& 2.49
		& 188.1		& 30.53
		& $\ldots$	& $\ldots$
		& $\ldots$	& $\ldots$
		& $10^{17.30}$	& $1.8$				\\
        B0540-69
		& 49.4
		& 124.7		& 31.00
		& $\ldots$	& $\ldots$
		& $\ldots$	& $\ldots$
		& $10^{14.15}$	& $2.0$				\\
        B1509-58
		& 4.40
		& 41.7		& 31.19
		& $\ldots$	& $\ldots$
		& $\ldots$	& $\ldots$
		& $10^{14.04}$	& $1.1$				\\
        J1617-5055	
		& 3.30
		& 90.6		& 30.78
		& $\ldots$	& $\ldots$
		& $\ldots$	& $\ldots$
		& $10^{12.64}$	& $1.6$				\\
        J0822-4300
		& 2.20
		& 83.4		& 30.53
		& 280		& 0.040
		& $\ldots$	& $\ldots$
		& $\ldots$	& $\ldots$			\\
        Vela	
		& 0.50
		& 61.3		& 30.53
		& 150		& 0.066
		& $\ldots$	& $\ldots$
		& $\ldots$	& $\ldots$			\\
        B1951+32
		& 2.5
		& 159		& 29.68
		& $\ldots$	& $\ldots$
		& $\ldots$	& $\ldots$
		& $10^{13.55}$	& 1.6				\\
        B1821-24
		& 5.1
		& 2060		& 27.35
		& $\ldots$	& $\ldots$
		& $\ldots$	& $\ldots$
		& $10^{16.36}$	& $1.2$				\\
        B0656+14
		& 0.76
		& 15.3		& 30.67
		& 67		& 4.5
		& 129		& $10^{-1.49}$
		& $10^{5.25}$	& $1.5$				\\
        Geminga
		& 0.16
		& 26.5		& 30.21
		& 48		& 0.16
		& $\ldots$	& $\ldots$
		& $10^{5.00}$	& $1.6$				\\
        B1055-52
		& 1.53
		& 31.9		& 30.03
		& 68		& 7.3
		& 320		& $10^{-3.64}$
		& $\ldots$	& $\ldots$			\\
        J0437-4715
		& 0.180
		& 1092		& 26.50
		& 22		& 0.16
		& 95		& $10^{-3.28}$
		& $\ldots$	& $\ldots$			\\
        \hline
      \end{tabular}
    \end{minipage}
\end{table*}

\noindent
{\bf Crab} \quad
From HEAO 1 observations,
its X-ray field is expressed by a power-law with
$\alpha=-1.81 \pm 0.02$ in the primary pulse (P1) phase (Knight 1982).
From the nebula and background subtracted counting rates,
its normalization factor of this power-law emission
becomes $N_{\rm pl}=5.3 \times 10^{15}d{}^2 (r_0/\rlc)^{-2}$,
where $d$ refers to the distance in kpc.  \\
{\bf B0540-69} \quad
From ASCA observations in 2-10 keV band, 
its X-ray radiation is known to be well fitted by
a power-law with $\alpha=-2.0$.
The unabsorbed luminosity in this energy range is 
$1.3 \times 10^{36} \mbox{erg s}{}^{-1}$,
which leads to $N_{\rm pl}= 9.0 \times 10^9 d^2 (r_0/\rlc)^{-2}$
(Saito 1998).\\
{\bf B1509-58} \quad
From ASCA observations in 2-10 keV band, 
its pulsed emission can be fitted by
a power-law with $\alpha=-1.1$.
The unabsorbed flux in this energy range is 
$2.9 \times 10^{-11} \mbox{erg/cm}^2 / \mbox{s}$,
which leads to $N_{\rm pl}= 9.3 \times 10^{11} d^2 (r_0/\rlc)^{-2}$. \\
{\bf J1617-5055} and {\bf J0822-4300} \quad
These two pulsars have resemble parameters such as
$\Omega \sim 90 \,\mbox{rad s}^{-1}$,
$\mu \sim 10^{30.6} \mbox{G cm}^3$,
and characteristic age $\tau \sim 8 \times 10^3$yr.
From the ASCA observations of J1617-5055 in 3.5-10 keV band, 
its pulsed emission 
subtracted the background and the steady components
can be fitted by a power-law with $\alpha=-1.6$
(Torii et al. 1998).
Adopting the distance to be 3.3 kpc (Coswell et al. 1975),
we can calculate its unabsorbed flux as
$3.1 \times 10^{-12} \mbox{erg/cm}^2 / \mbox{s}$,
which yields $N_{\rm pl}= 6.5 \times 10^{10} d^2 (r_0/\rlc)^{-2}$. 
On the other hand, 
the distance of J0822-4300 was estimated from VLA observations
as $d=2.2 \pm 0.3$kpc (Reynoso et al. 1995). 
ROSAT observations revealed that the soft X-ray
emission of this pulsar is consistent with a single-temperature
blackbody model with $kT_{\rm s}=0.28\pm 0.10$keV
and $A_{\rm s} \sim 0.04 A_* (d/2.2)^2$
(Petre, Becker, \& Winkler 1996). \\
{\bf Vela} \quad
From ROSAT observations in 0.06-2.4 keV, 
the spectrum of its point source (presumably the pulsar) emission 
is expressed by two components:
Surface blackbody component with 
$kT_{\rm s}=150$eV and $A_{\rm s}=0.066 A_* (d/0.5)^2$,
and a power-law component with $\alpha=-3.3$.
However, the latter component does not show pulsations;
therefore, we consider only the former component
as the X-ray field illuminating the outer gap. \\
{\bf B1951+32} \quad
From ROSAT observations in 0.1-2.4 keV,
the spectrum of its point source (presumably the pulsar) emission 
can be fitted by a single power-law component with $\alpha=-1.6$ and
intrinsic luminosity of 
$2.3 \times 10^{33} (d/2.5)^2 \mbox{erg s}^{-1}$
(Safi-Harb \& $\ddot{\rm O}$gelman 1995),
which yields 
$N_{\rm pl}=9.1 \times 10^{11} d^2 (r_0/\rlc)^{-2}$.
The extension of this power-law is consistent with the upper limit 
of the pulsed component in 2-10 keV energy band
(Saito 1998). \\
{\bf B1821-24} \quad
From ASCA observations in 0.7-10 keV band, 
its pulsed emission subtracted the background and the steady components
can be fitted by a power-law with $\alpha=-1.2$
(Saito 1997).
The unabsorbed flux in this energy range is 
$3.5 \times 10^{-12} \mbox{erg cm}^{-2} \mbox{s}^{-1}$,
which leads to $N_{\rm pl}= 1.4 \times 10^{14} d^2 (r_0/\rlc)^{-2}$. \\
{\bf B0656+14} \quad
Combining ROSAT and ASCA data, Greiveldinger et al. (1996) reported 
that the X-ray spectrum consists of three components:
the soft surface blackbody with $kT_{\rm s}=67$ eV and 
$A_{\rm s}=4.5 A_* (d/0.76)^2$,
a hard blackbody with $kT_{\rm h}=129$ eV and 
$A_{\rm h}=3.2 \times 10^{-2} A_* (d/0.76)^2$,
and a power law with $\alpha=-1.5$ and
$N_{\rm pl} = 3.1 \times 10^{5} d^2$.
The hard blackbody component takes the major role in maintaining the gap,
by virtue of its large emitting area. \\
{\bf Geminga} \quad
The X-ray spectrum consists of two components:
the soft surface blackbody with $kT_{\rm s}=50$ eV and 
$A_{\rm s}= 0.22 A_* (d/0.16)^2$
and a hard power law with $\alpha= -1.6$ and
$N_{\rm pl} = 3.9 \times 10^6 d^2$ (Halpern \& Wang 1997). 
A parallax distance of 160pc was estimated from HST observations
(Caraveo et al. 1996). \\
{\bf B1055-52} \quad 
Combining ROSAT and ASCA data, Greiveldinger et al. (1996)
reported that the X-ray spectrum consists of two components:
a soft blackbody with $kT_{\rm s}=68$ eV and 
$A_{\rm s}= 7.3 A_* (d/1.53)^2$
and a hard blackbody with $kT_{\rm h}=320$ eV and 
$A_{\rm h}= 2.3 \times 10^{-4} A_* (d/1.53)^2$. \\
{\bf J0437-4715} \quad
Using ROSAT and EUVE data 
(Becker \& Tr$\ddot{\rm u}$mper 1993; Halpern et at. 1996),
Zavlin and Pavlov (1998) demonstrated that both the spectra and the
light curves of its soft X-ray radiation can originate
from hot polar caps with a nonuniform temperature distribution and 
be modeled by a step-like functions having two different temperatures.
The first component is the emission from heated polar-cap core
with temperature $kT_{\rm h} = 10^{6.16}$ K measured at the
surface and with an area
$A_{\rm h} = 5.3 \times 10^{-4} A_* (d/0.180)^2$.
The second one can be interpreted as a cooler rim around the
polar cap on the neutron star surface with temperature
$kT_{\rm s} = 10^{5.53}$ K and with an area 
$A_{\rm s} = 0.16 A_* (d/0.180)^2$.
Considering the gravitational redshift factor of 0.76,
the best-fit temperatures observed at infinity become
$kT_{\rm s}= 22$ eV and $kT_{\rm h}= 95$ eV.
From parallax measurements, its distance is reported to be
$178 \pm 26$ pc (Sandhu et al. 1997).
We adopt $d=180$ pc as a representative value.

\subsection{Input Infrared Field}
\label{sec:input_IR}

We next consider the infrared photon field illuminating the gap.
In addition to the references cited below,
see also Thompson et al. (1999)
for Crab, B1509-58, Vela, B1951+32, Geminga, and B1055-52.

\begin{figure} 
\centerline{ \epsfxsize=8.5cm \epsfbox[200 2 700 350]{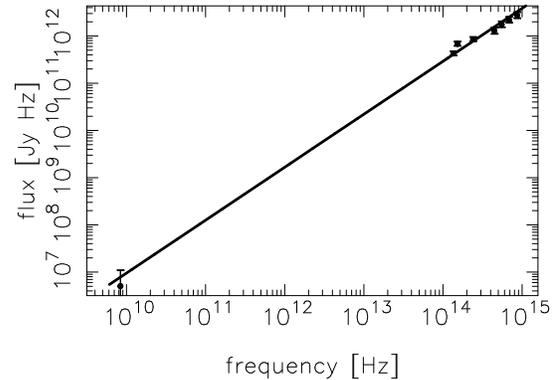} } 
\caption{\label{fig:spc_Crab} 
A single power-law fit of phase-averaged color spectrum
of the Crab pulsar.
The abscissa is the photon frequency in Hz,
while the ordinate is the photon flux in Jy$\cdot$Hz.
        }
\end{figure}

{\bf Crab} \quad
Interpolating the phase-averaged color spectrum in 
UV, U, B, V, R (Percival et al. 1993),
J, H, K (Eikenberry et al. 1997) bands,
and the radio observation at 8.4 GHz (Moffett and Hankins 1996),
the spectrum in IR energy range can be fitted by a
single power-law (fig.~\ref{fig:spc_Crab})
\begin{equation}
  \nu F_\nu = 5.75 \times 10^{-5} \nu^{1.12} \mbox{Jy Hz},
  \label{eq:spc_Crab}
\end{equation}
where $\nu$ is in Hz.
The Crab pulsar's flux at 8.4 GHz is, in fact, very uncertain, 
because it can scintillate away from Earth for tens of minutes
to hours.  
Moffett obtained a value of 0.61 mJy 
from the profiles he collected, while
Frail got a value of $0.5 \pm 0.1$ mJy from VLA imaging
(Moffett 2000, private communication).
They are within one sigma (0.1 mJy) of each other.
To estimate a conservative infrared photon number,
we simply assume that the flux at 8.4 GHz is 
$0.6 \pm 0.2$ mJy as a crude estimate,
because Moffett (1996) previously gave the value of 0.76 mJy.
If the error bar at 8.4 GHz is smaller,
then the infrared photon density around 0.01 eV reduces further;
this in turn results in a less upscattered, TeV flux.
Equation (\ref{eq:spc_Crab}) gives the following infrared number spectrum:
\begin{equation}
  dN_{\rm IR}/d\epsilon_{\rm IR}
    = 1.5 \times 10^{17} d^2 (r_0/\rlc)^{-2}
      \epsilon_{\rm IR}{}^{-0.88},
  \label{eq:IR_Crab}
\end{equation}
where $\epsilon_{\rm IR} m^{\rm e}c^2$ 
refers to the infrared photon energy.

{\bf B0540-69} \quad
Its de-extincted optical and soft X-ray pulsed flux densities 
can be interpolated as $F_\nu = 0.207 \nu^{-0.3}$ Jy
(Middleditch \& Pennypacker 1985). 
This line extrapolates to 0.47 mJy at 640 MHz,
which is consistent with an observed value 0.4 mJy
(Manchester et al. 1993).
We thus extrapolates the relation $F_\nu = 0.207 \nu^{-0.3}$
to the infrared energies and obtain 
\begin{equation}
  dN_{\rm IR}/d\epsilon_{\rm IR}
    = 1.6 \times 10^{12} d^2 (r_0/\rlc)^{-2}
      \epsilon_{\rm IR}{}^{-1.3}
  \label{eq:IR_B0540}
\end{equation}

{\bf B1509-58} \quad
The flux densities emitted from close to the neutron star
in radio (Taylor, Manchester, \& Lyne 1993), 
optical (Caraveo, Mereghetti, \& Bignami 1994),
soft X-ray (Seward et al. 1984),
and hard X-ray (Kawai et al. 1993) bands
can be fitted by a single power-law
$F_\nu = 1.36 \nu^{-0.32}$ Jy.
We thus adopt 
\begin{equation}
  dN_{\rm IR}/d\epsilon_{\rm IR}
    = 4.7 \times 10^{11} d^2 (r_0/\rlc)^{-2}
      \epsilon_{\rm IR}{}^{-1.3}
  \label{eq:IR_B1509}
\end{equation}

{\bf Vela} \quad
The flux densities emitted from close to the neutron star
in radio 
(Taylor, Manchester, \& Lyne 1993; Downs, Reichley, \& Morris 1973), 
optical (Manchester et al. 1980) bands
can be fitted by $F_\nu = 1.71 \times 10^7 \nu^{-0.91}$ Jy.
We thus adopt 
\begin{equation}
  dN_{\rm IR}/d\epsilon_{\rm IR}
    = 1.8 \times 10^{7} d^2 (r_0/\rlc)^{-2}
      \epsilon_{\rm IR}{}^{-1.9}
  \label{eq:IR_Vela}
\end{equation}

{\bf B1951+32},  \quad
The flux densities emitted from close to the neutron star
in radio (Taylor, Manchester, \& Lyne 1993)
and soft X-ray 
(Safi-Harb \& $\ddot{\rm O}$gelman, \& Finley 1995) bands
can be interpolated as $F_\nu = 32.8 \nu^{-0.49}$ Jy.
We thus adopt 
\begin{equation}
  dN_{\rm IR}/d\epsilon_{\rm IR}
    = 6.3 \times 10^{10} d^2 (r_0/\rlc)^{-2}
      \epsilon_{\rm IR}{}^{-1.5}
  \label{eq:IR_B1951}
\end{equation}

{\bf B0656+14} \quad
The nonthermal (most likely, magnetospheric) emission 
in I, R, V, B bands shows spectral index $-1.53$
(Caraveo et al. 1994a; Kurt et al. 1998),
which is much softer than that of Crab (eq.[\ref{eq:IR_Crab}]).
If we were to extrapolate it to the radio frequency,
the flux density at 1 GHz would exceed 100 Jy;
therefore, it is likely that the soft power-law spectrum
becomes harder below a cirtain frequency.
We thus estimate the upper limit of infrared flux density
by interpolating between radio 
(6 and 4 mJy at 0.4 and 1.4 GHz, respectively,
 Taylor, Manchester, \& Lyne 1993)
and the infrared-optical observations
($0.71$, $0.47$, $0.40$ $\mu$Jy at I, R, V bands).
The result is
\begin{equation}
  dN_{\rm IR}/d\epsilon_{\rm IR}
    = 2.0 \times 10^{1} d^2 (r_0/\rlc)^{-2}
      \epsilon_{\rm IR}{}^{-2.7}.
  \label{eq:IR_B0656}
\end{equation}

{\bf Geminga} \quad
The upper limit of the flux density in radio band 
(Taylor, Manchester, \& Lyne 1993)
and the flux density in optical band 
(Shearer et al. 1998)
gives spectral index greater (or harder) than $-0.69$.
Interpolating the infrared flux with these two frequencies,
we obtain 
\begin{equation}
  dN_{\rm IR}/d\epsilon_{\rm IR}
    = 1.9 \times 10^{7} d^2 (r_0/\rlc)^{-2}
      \epsilon_{\rm IR}{}^{-1.7},
  \label{eq:IR_Geminga}
\end{equation}
which gives a conservative upper limit of infrared photon number density
under the assumption of a single power-law interpolation.

{\bf B1055-52} \quad 
The flux densities emitted from close to the neutron star
in radio (Taylor, Manchester, \& Lyne 1993) 
and optical 
(Mignani et al. 1997) bands
can be interpolated as $F_\nu = 4.2 \times 10^5 \nu^{-0.77}$ Jy.
We thus adopt 
\begin{equation}
  dN_{\rm IR}/d\epsilon_{\rm IR}
    = 7.7 \times 10^{7} d^2 (r_0/\rlc)^{-2}
      \epsilon_{\rm IR}{}^{-1.8},
  \label{eq:IR_B1055}
\end{equation}

{\bf J1617-5055}, {\bf J0822-4300}, {\bf B1821-24}, 
and {\bf J0437-4715} \quad
There have been no available infrarad or optical observations
for these four pulsars.
We thus simply assume that $\alpha=-1.5$ for these four pulsars
and that $\nu F_\nu = 10^9$ Jy$\cdot$Hz at 0.01 eV.
We then obtain
\begin{equation}
  dN_{\rm IR}/d\epsilon_{\rm IR}
    = 2.5 \times 10^{7} d^2 (r_0/\rlc)^{-2}
      \epsilon_{\rm IR}{}^{-1.5}
  \label{eq:IR_J1617}
\end{equation}
for J1617-5055, while 
\begin{equation}
  dN_{\rm IR}/d\epsilon_{\rm IR}
    = 2.1 \times 10^{7} d^2 (r_0/\rlc)^{-2}
      \epsilon_{\rm IR}{}^{-1.5},
  \label{eq:IR_J0822}
\end{equation}
for J0822-4300, 
\begin{equation}
  dN_{\rm IR}/d\epsilon_{\rm IR}
    = 1.3 \times 10^{14} d^2 (r_0/\rlc)^{-2}
     \epsilon_{\rm IR}{}^{-1.5},
  \label{eq:IR_B1821}
\end{equation}
for B1821-24, and
\begin{equation}
  dN_{\rm IR}/d\epsilon_{\rm IR}
    = 3.6 \times 10^{13} d^2 (r_0/\rlc)^{-2}
      \epsilon_{\rm IR}{}^{-1.5},
  \label{eq:IR_J0437}
\end{equation}
for J0437-4715.

\subsection{Curvature-radiated $\gamma$-rays}
\label{sec:gamma-ray}

In this subsection, we present the results of GeV emission
via curvature process
and compare them with observations.
The results are presented for the two assumed values of
$\alpha= 30^\circ$ and $45^\circ$. \\

\noindent
\begin{table*}
  \centering
    \begin{minipage}{140mm}
      \caption{Expected $\gamma$-ray properties}
      \begin{tabular}{@{}lcccccccc@{}}
        \hline
        \hline
        pulsar	
		& $\inc$
		& $l_{\rm acc} / H$
		& $H / \rlc$
		& $E_{\rm c}$		
		& $L_{\rm GeV}$
		& $L_{\rm GeV}/d^2$
		& $\nu_{\rm peak}$
		& $(\nu F_\nu)_{\rm TeV}{}^\ddagger$	 \\
        \	
		& deg
		& 
		&
		& GeV			
		& $\mbox{erg s}{}^{-1}$
		& Jy$\cdot$Hz
		& Hz
		& Jy$\cdot$Hz				\\
        \hline
        Crab	
		& 30
		& 0.82
		& 0.023
		& 4.2
		& $1.3 \cdot 10^{33}$
		& $2.1 \cdot 10^{12}$
		& $2.5 \cdot 10^{27}$
		& $1.9 \cdot 10^{12}$				\\
	\ 	& 45
		& 0.96
		& 0.016
		& 5.2
		& $1.8 \cdot 10^{33}$
		& $3.0 \cdot 10^{12}$
		& $2.7 \cdot 10^{27}$
		& $4.4 \cdot 10^{12}$				\\
        B0540-69
		& 30
		& 0.21
		& 0.048
		& 7.8
		& $8.5 \cdot 10^{33}$
		& $3.6 \cdot 10^{10}$
		& $2.7 \cdot 10^{27}$
		& $8.3 \cdot 10^{10}$				\\
	\ 	& 45
		& 0.24
		& 0.034
		& 9.8
		& $1.3 \cdot 10^{34}$
		& $5.3 \cdot 10^{10}$
		& $2.9 \cdot 10^{27}$
		& $2.1 \cdot 10^{11}$				\\
        B1509-58
		& 30
		& 0.21
		& 0.075
		& 5.1
		& $5.7 \cdot 10^{33}$
		& $3.1 \cdot 10^{12}$
		& $3.4 \cdot 10^{27}$
		& $1.1 \cdot 10^{11}$				\\
	\ 	& 45
		& 0.23
		& 0.053
		& 6.6
		& $9.0 \cdot 10^{33}$
		& $4.9 \cdot 10^{12}$
		& $3.6 \cdot 10^{27}$
		& $2.9 \cdot 10^{11}$				\\
        J1617-5055
		& 30
		& 0.093
		& 0.097
		& 8.7
		& $1.4 \cdot 10^{34}$
		& $7.4 \cdot 10^{12}$
		& $1.2 \cdot 10^{27}$
		& $1.0 \cdot 10^{8}$				\\
	\ 	& 45
		& 0.10
		& 0.068
		& 11.1
		& $2.2 \cdot 10^{34}$
		& $1.1 \cdot 10^{13}$
		& $2.0 \cdot 10^{27}$
		& $2.5 \cdot 10^{8}$				\\
        J0822-4300
		& 30
		& 0.69
		& 0.045
		& 2.5
		& $5.7 \cdot 10^{32}$
		& $1.2 \cdot 10^{12}$
		& $6.0 \cdot 10^{26}$
		& $1.0 \cdot 10^{7}$				\\
	\ 	& 45
		& 0.57
		& 0.035
		& 3.9
		& $1.4 \cdot 10^{33}$
		& $3.1 \cdot 10^{12}$
		& $7.6 \cdot 10^{26}$
		& $3.7 \cdot 10^{7}$				\\
        Vela	
		& 30
		& 0.27
		& 0.084
		& 3.5
		& $1.7 \cdot 10^{33}$
		& $7.0 \cdot 10^{13}$
		& $3.2 \cdot 10^{26}$
		& $8.4 \cdot 10^{10}$				\\
	\ 	& 45
		& 0.26
		& 0.062
		& 4.9
		& $3.3 \cdot 10^{33}$
		& $1.4 \cdot 10^{14}$
		& $4.0 \cdot 10^{26}$
		& $2.6 \cdot 10^{11}$				\\
        B1951+32
		& 30
		& 0.16
		& 0.092
		& 5.4
		& $2.8 \cdot 10^{33}$
		& $4.7 \cdot 10^{12}$
		& $6.3 \cdot 10^{26}$
		& $3.1 \cdot 10^{10}$				\\
	\ 	& 45
		& 0.18
		& 0.064
		& 6.9
		& $4.2 \cdot 10^{33}$
		& $7.1 \cdot 10^{12}$
		& $7.4 \cdot 10^{26}$
		& $8.0 \cdot 10^{10}$				\\
        B1821-24
		& 30
		& 0.22
		& 0.067
		& 5.2
		& $4.6 \cdot 10^{32}$
		& $1.8 \cdot 10^{11}$
		& $5.0 \cdot 10^{26}$
		& $6.2 \cdot 10^{7}$				\\
	\ 	& 45
		& 0.25
		& 0.047
		& 6.7
		& $7.2 \cdot 10^{32}$
		& $2.9 \cdot 10^{11}$
		& $4.7 \cdot 10^{26}$
		& $1.6 \cdot 10^{8}$				\\
        B0656+14
		& 30
		& 0.43
		& 0.160
		& 1.1
		& $2.0 \cdot 10^{32}$
		& $3.7 \cdot 10^{12}$
		& $3.7 \cdot 10^{26}$
		& $1.7 \cdot 10^{12}$				\\
	\ 	& 45
		& 0.69
		& 0.098
		& 1.2
		& $1.8 \cdot 10^{32}$
		& $3.2 \cdot 10^{12}$
		& $3.8 \cdot 10^{26}$
		& $3.0 \cdot 10^{12}$				\\
        Geminga	
		& 30
		& 0.059
		& 0.334
		& 4.0
		& $4.1 \cdot 10^{33}$
		& $1.7 \cdot 10^{15}$
		& $1.5 \cdot 10^{27}$
		& $7.0 \cdot 10^{9}$				\\
	\ 	& 45
		& 0.083
		& 0.213
		& 4.5
		& $4.3 \cdot 10^{33}$
		& $1.8 \cdot 10^{15}$
		& $6.5 \cdot 10^{26}$
		& $1.4 \cdot 10^{10}$				\\
        B1055-52
		& 30
		& 0.62
		& 0.130
		& 0.98
		& $8.5 \cdot 10^{31}$
		& $3.8 \cdot 10^{11}$
		& $6.0 \cdot 10^{26}$
		& $6.3 \cdot 10^{9}$				\\
	\ 	& 45
		& $1.0{}^\dagger$
		& 0.078
		& 1.0
		& $7.1 \cdot 10^{31}$
		& $3.2 \cdot 10^{11}$
		& $6.0 \cdot 10^{26}$
		& $1.1 \cdot 10^{10}$				\\
        J0437-4715
		& 30
		& 0.043
		& 0.371
		& 5.0
		& $6.1 \cdot 10^{32}$
		& $2.0 \cdot 10^{14}$
		& $6.3 \cdot 10^{26}$
		& $5.1 \cdot 10^{7}$				\\
	\ 	& 45
		& 0.062
		& 0.233
		& 5.4
		& $6.1 \cdot 10^{32}$
		& $2.0 \cdot 10^{14}$
		& $5.9 \cdot 10^{26}$
		& $9.9 \cdot 10^{7}$				\\
      \hline
      \end{tabular}
      \begin{flushleft}
        ${}^\dagger$ 
          The Lorentz factors are assumed to be
          the value that can be attained if particles are
          accelerated by $\bar{\Ell}$ in length $H$.\\
        ${}^\ddagger$ 
          The TeV flux are evaluated at the $\nu F_\nu$ peak frequency,
          $\nu_{\rm peak}$.
      \end{flushleft}
    \end{minipage}
\end{table*}

For Crab, B0540-69, B1509-58, and J1617-5055,
the X-ray field is dominated by the power-law component
to have high number densities above $10^{15} \mbox{cm}^{-3}$.
In this case, the gap half widths are less than $10\%$ of $\rlc$.
The intrinsic luminosities of these young pulsars 
in GeV energy range exceed $10^{33} \mbox{ergs s}^{-1}$.
Except for the distant pulsar B0540-69, their GeV fluxes
are expected to be large enough to be observed with 
a space $\gamma$-ray telescope.

For the relatively young pulsars J0822-4300 and Vela,
the X-ray field is dominated by the surface blackbody component;
the number density ($N_{\rm x}$) becomes about 
$10^{14} \mbox{cm}^{-3}$.
The gap half width is about $5\%$ of $\rlc$ and the intrinsic GeV luminosity
is $\sim 10^{33} \mbox{ergs s}^{-1}$.

For the middle-aged pulsar B1951+32,
its relatively strong magnetic field at the gap center
($B \sim 10^6$G)
results in a strong GeV emission like the young pulsars.

The millisecond pulsar B1821-24 has a very strong magnetic field
($B \sim 10^{7.5}$G) at the gap center.
However, its strong ($N_{\rm x} \sim 10^{17.5}\mbox{cm}^{-3}$)
X-ray field prevents the gap to extend in the magnetosphere.
As a result, $L_{\rm GeV}$ is relatively small compared with other
pulsars.

For the three middle-aged pulsars B0656+14, Geminga, and B1055-52,
their power-law components are too weak to dominate
the surface blackbody emissions;
the surface emissions are also weak 
($N_{\rm X} < 10^{14.2}$) to allow the gaps to extend
more than $10\%$ of $\rlc$.
However, the extended gaps do not mean large 
intrinsic GeV luminosities, 
because their small magnetic fields ($B < 10^{4.5}$G)
suppress the acceleration field.
In the case of Geminga, its proximity leads to a large GeV flux.

In the case of the millisecond pulsar J0437-4715,
its weak X-ray field ($N_{\rm x} < 10^{14} \mbox{cm}^{-3}$)
due to the soft and hard blackbody emissions
results in an extended gap.
This active gap, together with its proximity,
leads to a large GeV flux next to Geminga.
However, its relatively strong magnetic field ($B \sim 10^6$ G)
at the gap center may indicate the presence of an additional
power-law component,
which reduces the gap width and hence the GeV flux.
Therefore, further hard X-ray observations are necessary
for this millisecond pulsar.

\subsection{Invisibility of TeV pulses}
\label{sec:invisib}

If an electron or a positron is migrating 
with Lorentz factor $\Gamma \gg 1$ in an isotropic photon field,
it upscatters the soft photons to produce
the following number spectrum of $\gamma$-rays
(Blumenthal \& Gould 1970):
\begin{eqnarray}
  \frac{d^2 N}{dtd\Eg}
  &=& \frac34 \sgT \frac{c}{\Gamma^2}
      \frac{dN_{\rm IR}}{d\epsilon_{\rm IR}}
      \frac{d\epsilon_{\rm IR}}{\epsilon_{\rm IR}}
  \nonumber \\
  & & \hspace{-2.0 truecm}
      \times
      \left[ 2q \ln q +(1+2q)(1-q)
            +\frac{(Qq)^2(1-q)}{2(1+Qq)}
      \right],
  \label{eq:spc_tev}
\end{eqnarray}
where $Q \equiv 4 \epsilon_{\rm IR} \Gamma$ and 
$q \equiv \Eg / Q(\Gamma-\Eg)$;
here, $\Eg$ refers to the energy of the upscattered photons in 
$m_{\rm e}c^2$ unit.
Substituting the infrared photon spectrum $dN_{\rm IR}/d\epsilon_{\rm IR}$,
integrating $d^2 N/dtd\Eg$ over $\epsilon_{\rm IR}$,
and multiplying the $\gamma$-ray energy ($\Eg m_{\rm e}c^2$) and
the electron number ($N_{\rm e}$) in the gap,
we obtain the flux density of the upscattered, TeV photons 
as a function of $\Eg$.

\begin{figure} 
\centerline{ \epsfxsize=8.5cm \epsfbox[200 2 700 350]{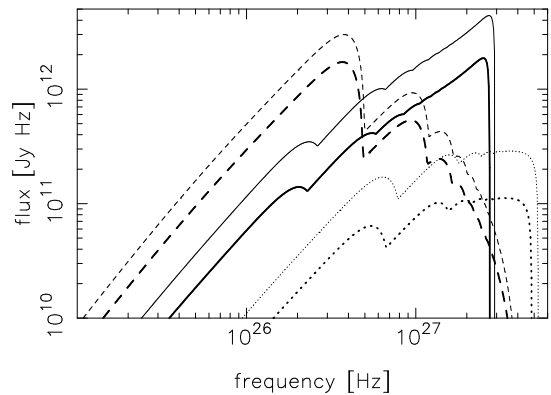} } 
\caption{\label{fig:spc_tev} 
Expected TeV spectra of the three brightest pulsars:
Crab (solid lines), B0656+14 (dashed ones), and B1509-58 (dotted ones).
The thick and thin curves represent the cases of 
$\inc= 30^\circ$ and $45^\circ$, respectively.
The abscissa is the photon frequency in Hz,
while the ordinate is the photon flux in Jy$\cdot$Hz.
        }
\end{figure}

We compute the TeV spectra of individual pulsars and summarize the
results in table~3;
the $\nu F_\nu$ peak frequencies and the fluxes are given in the
last two columns.
Moreover, for the three brightest pulsars
(Crab, B0656+14, B1509-58),
their computed $\nu F_\nu$ spectra are presented in figure~\ref{fig:spc_tev}.
It follows from table~3 and figure~\ref{fig:spc_tev}
that the TeV emissions are invisible with the current ground-based telescopes,
except for Crab and B0656+14.
Since $H$ becomes as small as $l_{\rm acc}$ for the Crab pulsar,
significant fraction of the particles are, in fact, 
unsaturated.
Therefore, the expected TeV flux is overestimated.
To constrain the absolute pulsed TeV flux from the Crab pulsar,
we must discard the mono-energetic approximation for the particle
distribution function and explicitly solve the Boltzmann equations of
particles and $\gamma$-rays, together with the Poisson equation for $\Phi$,
under suitable boundary conditions.
In addition, considering the fact that the tertiary infrared photons
are not isotropic but have small collision angles with the particles,
we can understand that the TeV flux computed from 
equation~(\ref{eq:spc_tev}) are, in general, overestimated.
For B1055-52 and B0656+14,
the emitting areas of the soft blackbody components are unnaturally large.
Therefore, more accurate X-ray observations are necessary 
for a quantitative prediction of their TeV fluxes.

It is worth noting that the TeV fluxes are less than $3 \%$ of
the GeV fluxes except for Crab, B0540-69, and B0656+14;
this conclusion is qualitatively consistent with Romani (1996).
The predicted TeV flux from B0540-69 is greater than the GeV flux,
because (for $\inc=30^\circ$ for instance)
its infrared photon number density in the energy interval 0.01 eV and 0.1 eV
attains $7.5 \times 10^{18} \mbox{cm}^{-3}$, 
which well exceeds the X-ray number density
$6.4 \times 10^{17} \mbox{cm}^{-3}$ between 0.1 keV and 100 keV.

\section{Discussion}
\label{sec:discussion}

\subsection{Summary}
\label{sec:summary}
To sum up, we have considered the electrodynamic structure of 
an outer gap accelerator in which relativistic particles
emit $\gamma$-rays via curvature process.
Imposing the gap closure condition that a single pair produces
one pair in the gap on average,
we solve selfconsistently the gap width as a function of 
the X-ray fields and the pulsar parameters.
Once the gap width is known, we can further compute
the acceleration field and the resultant $\gamma$-ray emissions.
It was demonstrated that 
the luminosities of GeV and TeV emissions 
are a decreasing function of the X-ray energy and number density.
We also showed that the expected $\nu F_\nu$ fluxes
($<10^{11.5}$Jy~Hz) of IC-scattered, TeV $\gamma$-rays from 
the outer gaps of rotation-powered pulsars
are less than the observational upper limits,
except for Crab and B0656+14.
For Crab, energy-dependent particle distribution function should be
considered, whereas 
for B0656+14, more accurate X-ray observations are required.
It is concluded that 
the difficulty of excessive TeV emission,
which appears in the CHR picture,
does not arise in the present outer gap model.

\subsection{Stability of the Gap}
\label{sec:stability}
The outer gap in the present model is stable,
regardless of whether the X-ray field is dominated by 
a surface blackbody or a magnetospheric power-law component.
Consider the case when the gap width $H$ slightly increases
as an initial perturbation.
It increases both $\Ell$ and $V_{\rm gap}$, 
which in turn increases both $E_{\rm c}$ and $N_\gamma$.
The increase of $E_{\rm c}$ results in the decrease of $E_{\rm th}$. \\
{\bf (1)} \quad When the surface blackbody dominates the X-ray field,
the X-ray spectrum and luminosity are unchanged by the perturbation.
Therefore, the decrease of $E_{\rm th}$ implies the decrease of 
$\lambda_1$ or $\lambda_2$ and hence $\lambda_{\rm p}$. \\
{\bf (2)} \quad When the magnetospheric emission dominates,
the secondary and tertiary emissions will increase with 
$E_{\rm c}$ and $N_\gamma$; 
therefore, $N_{\rm pl}$ increases as well.
Accordingly, the decrease of $E_{\rm th}$ and the increase of
$N_{\rm pl}$ imply a significant decrease of $\lambda_3$ and
hence $\lambda_{\rm p}$. \\
In either case, it follows that $\lambda_{\rm p}$ decreases 
owing to the initial increase of $H$.
Reminding the gap closure condition $H=\lambda_{\rm p}/N_\gamma$,
we find a negative feedback which cancels the initial perturbation
of $H$.

\subsection{Pulse Sharpness}
\label{sec:sharpness}
Let us discuss the expected sharpness of GeV pulses.
It seems unlikely that the azimuthal width of the gap increases
with decreasing $H$.
Therefore, it would be possible to argue that the solid angle
in which the primary $\gamma$-rays are emitted decreases as the
arc of the gap along the last open field line (i.e., $2H$) decreases.
On these grounds, we can expect a sharp pulse 
when $h \ll 1$ holds, such as for Crab and J0822-4300.

Qualitatively speaking, the same conclusion can be expected for
millisecond pulsars and magnetars.
In the case of a millisecond pulsar,
its fast rotation shrinks the light cylinder.
In such a small-volume magnetosphere, 
the outer gap is immersed in a dense magnetospheric,
power-law X-ray emission.
As a result, $\lambda_{\rm p}$ decreases to reduce $h$.
In the case of a magnetar,
its strong magnetic field makes the expansion coefficient $A$
in equation (\ref{eq:Poisson-2}) be large.
Therefore, a very thin ($h \ll 1$) gap with a strong $\Ell$
would be expected.
In short, for young pulsars, millisecond pulsars, and magnetars,
their high-energy pulsations are expected to show sharp peaks.

\subsection{Validity of Assumptions}
\label{sec:valid_assump}
First, we reduced the Poisson equation 
into the one-dimensional form (eq.[\ref{eq:Poisson-2}]),
by assuming $D_\perp \gg H$. 
Let us briefly consider the two-dimensional effect due to
the transfield derivative in the Poisson equation.
When $D_\perp$ becomes small,
the gap shifts outwards, $\Ell$ is partially screened, and $H$ enlarges 
(fig.~12 in Paper I; see also Cheng, Ho, \& Rudermann 1986a
 for a screened, or spatially constant $\Ell$ in a thin gap).
Owing to the screened acceleration field,
the GeV and TeV fluxes becomes small
compared with those obtained in $D_\perp \gg H$ case.
On these grounds, we can constrain the upper limit of the TeV fluxes 
in the transversely thick limit, $D_\perp \gg H$.''

Secondly, let us discuss the case 
when the assumption of the vacuum gap breaks down.
In this case, the charges in the gap
partially cancel the original $\Ell$ obtained in the vacuum gap
(eq. [\ref{eq:Poisson-2}]).
The partially screened $\Ell$ results in the
decrease of the TeV fluxes.
On these grounds, we can regard the TeV fluxes
presented in the present paper as the firm upper limits.

Thirdly, we consider the influence of cyclotron resonance scatterings.
For one thing, the {\it soft} blackbody emission from the whole surface
may be scattered to be anisotropic (Daugherty \& Harding 1989).
Such effects are important for polar cap models, because
the collision angles ($\cos^{-1}\mu_{\rm c}$) 
suffer significant corrections.
Nevertheless, in an outer gap, such corrections are negligibly small.
Moreover, the cyclotron resonance increases the
effective emitting area and decreases the temperature.
For simplicity, we neglect these two effects in this paper,
because they cancel each other.
For example, the decreased temperature results in a decrease of the 
target photons above a certain threshold energy for pair production.
On the other hand, the increased emitting area increases the number
of target photons above the threshold,
thereby cancel the effect of the decreased temperature.
What is more, 
the {\it hard} blackbody emission from the heated polar caps
may be scattered to be smeared out.
That is, most of the hard X-rays may be scattered back to 
the stellar surface owing to cyclotron resonance scatterings 
and reemitted as soft X-rays (Halpern \& Ruderman 1993).
In this case, the hard component will be indistinguishable
with the original soft component due to the neutron-star cooling.
Nevertheless, for older pulsars such as 
B0656+14 and B1055-52,
these effect seems to be ineffective 
probably due to their less dense electrons around the polar cap
near the neutron star surface.

\subsection{Gamma-ray Luminosity vs. Spin-down Luminosity}
\label{sec:Lgev-Lspin}
Curvature-radiated luminosity,
$L_{\rm GeV}$, has a weak dependence on the spin-down
luminosity, $L_{\rm spin}$, 
if we fix the transfield thickness of the gap, $D_\perp/\rlc$.
In another word, the evolution of $D_\perp/\rlc$ is crucial
to discuss the $L_{\rm GeV} \propto L_{\rm spin}{}^{0.5}$ relation
(Thompton et al. 1994; Nel et al. 1996).
To solve $D_\perp$, 
we must analyze the two-dimensional Poisson equation 
on the poloidal plane;
however, it is out of the scope of the present paper.

\subsection{Synchrotron Radiation Below 10 MeV Energies}
\label{sec:sycnrotron}
As we have seen, the accelerated particles reach curvature-radiation
reaction limit to become roughly monoenergetic.  
The curvature spectrum in lower energies
then becomes a
power law with a spectral index $1/3$, 
which is much harder than the observed $\gamma$-ray pulsar spectra.
In this subsection, we demonstrate that the $\gamma$-ray spectrum
below a certain energy (say $10$ MeV) is dominated by 
a synchrotron radiation from freshly born particles
and that the expected $\gamma$-ray spectra further softens.

As an example exhibiting a soft power-law $\gamma$-ray spectrum
from eV to GeV energies,
we consider the Crab pulsar.
To discriminate whether curvature or synchrotron process dominates,
we separately consider each process and take the ratio of the 
radiation-reaction forces.
That is, we ignore much complicated synchro-curvature process,
because such details are not important for the present purpose.

Let us first consider the case of $\inc=45^\circ$,
which gives $B_5=6.5 \times 10^2$ at the gap center.
Since the curvature-radiated $\gamma$-ray energy is 
$E_{\rm c}=5.2$ GeV,
the freshly born particles have the
Lorentz factors of $\Gamma_0 \sim 5 \times 10^3$.
A particle with this Lorentz factor emit synchrotron radiation around 
the energy
\begin{equation}
  h \nu_{\rm sync}= \frac{3 h \Gamma_0{}^2 eB\sin\chi_{\rm p}}
                      {4\pi m_{\rm e} c},
  \label{eq:synchro_Ec}
\end{equation}
where $\chi_{\rm p}$ denotes the pitch angle of the particles.

\begin{figure} 
\centerline{ \epsfxsize=9cm \epsfbox[70 50 400 460]{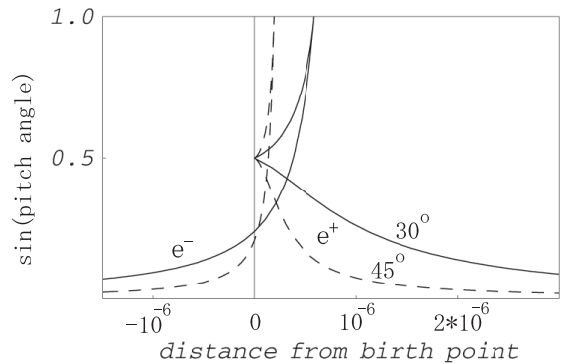} } 
\caption{\label{fig:pitch} 
Pitch angle evolution due to synchrotron radiation
as a function of the position.
The abscissa denotes the distance (in $\rlc$ unit) 
along the field lines from the birth place (distance$=0$).
The thick and thin lines correspond to $\inc=30^\circ$ 
and $\inc=45^\circ$, respectively. 
        }
\end{figure}

\begin{figure} 
\centerline{ \epsfxsize=9cm \epsfbox[70 50 400 460]{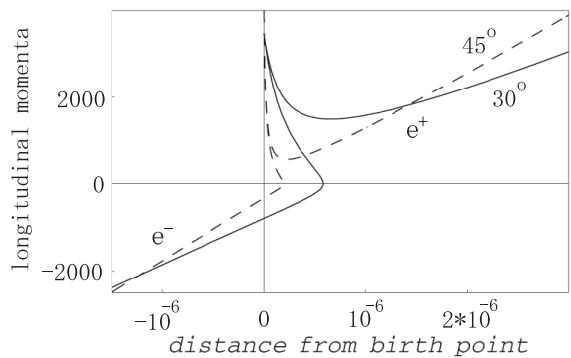} } 
\caption{\label{fig:synch} 
Longitudinal momentum evolution due to synchrotron radiation
as a function of the position.
The ordinate are nomalized in $m_{\rm e}c^2$ unit.
The abscissa and the lines are the same as figure~7.
        }
\end{figure}

We can solve the evolution of the Lorentz factor
and $\chi_{\rm p}$ 
simultaneously by the method described in \S~5.3 of Paper~I.
We present the evolution of $\sin\chi_{\rm p}$
due to synchrotron radiation shortly after the pair production
in figure~\ref{fig:pitch},
and the evolution of the longitudinal momenta 
in figure~\ref{fig:synch}.
In both figures,
the abscissa designates the distance along the fieldlines 
in $\rlc$ unit with respect to the birth place
(distance$=0$). 
The particles are supposed to be created with positive momenta;
therefore, electrons turn back to have negative longitudinal momenta.
Positrons lose longitudinal momenta on the initial stage of acceleration, 
because the relativistic beaming effect causes 
the synchrotron-radiation-reaction force 
not only in the transverse but also in the longitudinal directions.

It follows from figure~\ref{fig:pitch}
that we can approximate $\sin\chi_0 \sim 0.3$
when the particles have not run $\Delta l \sim 1 \times 10^{-6}\rlc$ 
for $\inc=45^\circ$.
When $\chi_{\rm p}$ is kept around $0.3$, 
Lorentz factors are also in the same order of $\Gamma_0$.
Substituting $\sin\chi_{\rm p}=0.3$ 
and $\Gamma_0 \sim 5 \times 10^3$ into equation (\ref{eq:synchro_Ec}), 
we obtain $h \nu_{\rm sync} \sim 8.5$ MeV 
as the central energy of the synchrotron spectrum for $\inc=45^\circ$.
The fraction of the particles having $\sin\chi_{\rm p} \sim 0.3$ and
Lorentz factor $\sim \Gamma_0$ to the saturated particles 
then becomes
\begin{equation}
  \frac{\Delta N_{\rm e}}{N_{\rm e}}
    \sim \frac{\Delta l}{H}
    \sim 6 \times 10^{-5}.
  \label{eq:ratio_population}
\end{equation}

We can estimate the ratio between the synchrotron radiation 
from the freshly born particles and the curvature radiation 
from the saturated particles
as follows:
\begin{equation}
  R_{\rm sc}(\nu)= \frac{\Delta N_{\rm e}}{N_{\rm e}}
       \frac{dP_{\rm sync}/d\nu}{dP_{\rm curv}/d\nu},
  \label{ratio_forces}
\end{equation}
where 
\begin{equation}
  \frac{dP_{\rm sync}}{d\nu}
    \equiv \frac{\sqrt{3}e^3 B \sin\chi_0}{m_{\rm e}c^2}
           F \left(\frac{h\nu}{h\nu_{\rm sync}}\right),
  \label{eq:def_Psync}
\end{equation}
\begin{equation}
  \frac{dP_{\rm curv}}{d\nu}
    \equiv \frac{\sqrt{3}e^2 \Gamma}{R_{\rm c}}
           F \left(\frac{h\nu}{E_{\rm c}}\right),
  \label{eq:def_Pcurv}
\end{equation}
\begin{equation}
  F(x) \equiv x \int^\infty_x K_{5/3}(y)dy;
  \label{eq:def_F}
\end{equation}
$K_{5/3}$ refers to the modified Bessel function of 5/3 order;
$\Gamma$ in equation~(\ref{eq:def_Pcurv}) denotes the saturated Lorentz 
factor and becomes $2.4 \times 10^7$ for $45^\circ$ for Crab.
At the synchrotron peak energy, $0.29 h\nu_{\rm sync}=2.5$ MeV,
the ratio becomes $R_{\rm sc}=12$.

In the same manner we can consider the case of $\inc=30^\circ$.
In this case, we have $h\nu_{\rm sync} \sim 1.7$ MeV and
$\Delta N_{\rm e}/N_{\rm e} \sim 1 \times 10^{-4}$.
As a result, we obtain $R_{\rm sc}=11$ at $0.5$ MeV.

We can therefore conlude that the $\gamma$-ray spectrum below
certain energy ($\sim 10$ MeV)
is dominated by the synchrotron radiation from freshly born particles.

In addition, in the case of Crab, the unsaturated motion of particles
($l_{\rm acc} \sim H$) implies that the synchro-curvature radiation
from unsaturated particles are also important.
Therefore, the spectrum below GeV will become much softer compared
with the simple curvature spectrum with central energy 
$E_{\rm c}=5.2$ (or $4.2$) GeV for $\inc=45^\circ$ (or $30^\circ$).

\subsection{Comparison with Zhang and Cheng model}
\label{sec:cf_ZC}
Finally, we point out the difference between the present work
and Zhang and Cheng (1997);
they considered a gap closure condition
so that the curvature-radiated $\gamma$-ray energy may be
adjusted just above the threshold of pair production.
That is, they considered the $\gamma$-ray
energy to be about 
$E_{\gamma, {\rm ZC}} \equiv (m_{\rm e}c^2)^2 / E_{\rm x}$,
where $E_{\rm x}$ refers to the characteristic X-ray energy.
By equating $E_{\gamma, {\rm ZC}}$ with the central energy
of curvature radiation (eq. [\ref{eq:Ec}] in our notation),
they closed the equations.
When the soft (or hard) blackbody emission dominates,
$E_{\rm x}$ can be approximated by
$3 kT_{\rm s}$ (or $3 kT_{\rm h}$).

\begin{figure} 
\centerline{ \epsfxsize=8.5cm \epsfbox[200 2 700 350]{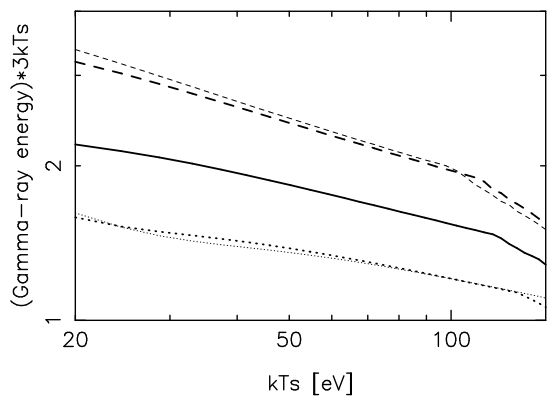} } 
\caption{\label{fig:cfZC} 
The ratio between $E_{\rm c}$ computed from equation (13)
and $E_{\gamma, {\rm ZC}}$ (see text).
Neither the hard blackbody nor the power-law components are
considered.
X-rays are supplied by the whole surface blackbody emission;
the abscissa refers to the temperature, $kT_{\rm s}$.
        }
\end{figure}

The model of Zhang and Cheng (1997) is, in fact, 
qualitatively consistent with our
gap closure condition, provided that the X-ray are supplied by
the soft or hard blackbody emission.
More specifically, our model gives about 2 times larger
characteristic $\gamma$-ray energy compared with their model.
To see this, we present in figure~\ref{fig:cfZC} 
the ratio between
$E_{\rm c}$ computed from equation (\ref{eq:Ec})
and $E_{\gamma, {\rm ZC}}$;
the hard blackbody or the power-law components are not considered
in this calculation.
The abscissa indicates the soft blackbody temperature, $kT_{\rm s}$. 
For the three thick curves, $\Omega_2$ is fixed at $0.5$;
the solid, dashed, and dotted lines corresponds to
$\mu_{30}= 1.0$, $3.0$, and $0.3$, respectively.
For the two thin curves, on the other hand,
$\mu_{30}$ is fixed at $1.0$;
the dashed and dotted curves corresponds to
$\Omega_2= 1.0$ and $0.25$, respectively.
At small $kT_{\rm s}$, 
our model gives more than twice greater $\gamma$-ray energy
compared with Zhang and Cheng (1997); nevertheless, 
the difference is not very prominent.

It should be noted, however, that 
the spectra of the X-ray radiation are explicitly considered 
in our present model 
in the sense we perform the integration over X-ray energies
in equations (\ref{eq:def_lambda_4}), (\ref{eq:def_lambda_5}),
and (\ref{eq:def_lambda_3})
and that the additional, power-law component is considered 
in our present model.

It would be interesting to investigate the back reaction 
of the accelerated particles on the polar cap heating,
which was deeply investigated by Zhang and Cheng (1997).
Consider the case when the voltage drop in the gap approaches 
the surface EMF, $V_*$.
Such an active gap will supply copious
relativistic primary particles
to heat up the polar cap due to bombardment.
The resultant hard blackbody emission supplies target 
photons for pair production to make the realistic
solution deviate from the thick solid curve 
and approach thin curves at small $kT_{\rm s}$ (fig.~\ref{fig:width}). 
Therefore, this sort of back reaction on the X-ray field 
due to the relativistic particles needs further consideration.

\acknowledgments

This research owes much to the helpful comments of Dr. S. Shibata.
The author wishes to express his gratitude to
Drs. Y. Saito and A. Harding for valuable advice. 
He also thanks the Astronomical Data Analysis Center of
National Astronomical Observatory, Japan for the use of workstations.

\end{document}